\documentclass[journal]{IEEEtran}
\IEEEoverridecommandlockouts
\usepackage{cite}
\usepackage{amsmath,amsfonts,amssymb,amsthm}
\usepackage{dsfont}
\usepackage{algorithmic}
\usepackage{graphicx}
\usepackage{textcomp}
\usepackage{xcolor}
\usepackage{hyperref}
\usepackage{todonotes}
\usepackage{algorithm}
\usepackage{tabularx,booktabs}
\usepackage{subcaption}
\usepackage{placeins}
\usepackage{soul} 

\newcolumntype{Y}{>{\centering\arraybackslash}X}
\usepackage{multirow}
\usepackage[font=small,labelfont=bf]{caption}

\usepackage{subfloat}

\usepackage{amsthm}\theoremstyle{plain}

\newtheorem{lemma}{Lemma}

\DeclareMathOperator*{\argmax}{argmax}

\def\BibTeX{{\rm B\kern-.05em{\sc i\kern-.025em b}\kern-.08em
    T\kern-.1667em\lower.7ex\hbox{E}\kern-.125emX}}

\newcommand{\rv}[1]{{\color{black} #1}}

\begin{document}

\title{Adversarial Node Placement in Decentralized Federated Learning: Maximum Spanning-Centrality Strategy and Performance Analysis
\thanks{
\IEEEauthorblockA{\IEEEauthorrefmark{1}Purdue University, Email: \{apiasecz, cgb\}@purdue.edu}
    
    \IEEEauthorblockA{\IEEEauthorrefmark{2}Princeton University, Email: er6214@princeton.edu}
    
    \IEEEauthorblockA{\IEEEauthorrefmark{3}
    Massachusetts Institute of Technology, Email: rohit100@mit.edu}

An abridged version of this work appeared in the IEEE International Conference on Communications (ICC)~\cite{piaseczny2024impact}.}
\thanks{This work is supported in part by the ONR under grants N000142212305 and N000142112472.}
}

\allowdisplaybreaks

\author{
    \IEEEauthorblockN{Adam Piaseczny\IEEEauthorrefmark{1}, Eric Ruzomberka\IEEEauthorrefmark{2}, Rohit Parasnis\IEEEauthorrefmark{3}, Christopher G. Brinton\IEEEauthorrefmark{1}}

}
\maketitle

\date{}
\begin{abstract}
As Federated Learning (FL) becomes more widespread, there is growing interest in its decentralized variants. Decentralized FL leverages the benefits of fast and energy-efficient device-to-device communications to obviate the need for a central server. However, this opens the door to new security vulnerabilities as well. While FL security has been a popular research topic, the role of adversarial node placement in decentralized FL remains largely unexplored. This paper addresses this gap by evaluating the impact of various coordinated adversarial node placement strategies on decentralized FL's model training performance. We adapt two threads of placement strategies to this context: maximum span-based algorithms, and network centrality-based approaches. Building on them, we propose a novel attack strategy, MaxSpAN-FL, which is a hybrid between these paradigms that adjusts node placement probabilistically based on network topology characteristics. Numerical experiments demonstrate that our attack consistently induces the largest degradation in decentralized FL models compared with baseline schemes across various network configurations and numbers of coordinating adversaries. We also provide theoretical support for why eigenvector centrality-based attacks are suboptimal in decentralized FL. Overall, our findings provide valuable insights into the vulnerabilities of decentralized FL systems, setting the stage for future research aimed at developing more secure and robust decentralized FL frameworks.

\end{abstract}

\section{Introduction}
\label{sec:introduction}
\noindent Federated Learning (FL) \cite{mcmahan2017communication} has emerged as a popular method for distributed machine learning at the network edge \cite{9084352}. By sharing the workload of the training process across different edge devices, FL enables parallel computation and eliminates the need for shared access to training data, enhancing both efficiency and privacy of the learning process.

Traditional FL frameworks follow a client-server architecture, where devices conduct local model updates and a server periodically aggregates them into a global model. Recent work \cite{9933813, raynal2023decentralized, 10251949} has investigated the fully decentralized FL setting, where devices execute the model aggregation step too, via iterative cooperative consensus formation over local wireless topologies formed through device-to-device (D2D) communication protocols. Decentralized FL eliminates the need for a central server, which can offer advantages in terms of communication efficiency and system scalability.

At the same time, the decentralized FL setting introduces novel security challenges. Like their centralized counterparts, decentralized FL systems are vulnerable to attacks on the training process by adversarial edge devices \cite{lyu2020threats, li2023backdoor, 9411833}. For instance, in poisoning attacks, adversaries maliciously interfere with training data samples and/or model updates at different devices, compromising the integrity of the model shared for aggregations (to the server in the centralized case, and to neighbors in the decentralized case), preventing the FL system from achieving its training target \cite{goodfellow2015explaining, huang2017adversarial, wang2023potent}. In the decentralized FL setting, however, the potency of attacks is not only affected by the methods adversaries employ to disrupt samples and model updates, but also by the \textit{location within the network} from which they launch their attacks. Their local topology impacts the number of nodes they can directly poison.

Motivated by this, in this paper, we aim to answer the following question: \textit{\textbf{How does the location or placement of adversarial nodes in the network affect the training process of decentralized FL systems?}} An answer to this question can provide valuable insights into the vulnerabilities of decentralized FL systems and lay the groundwork for future research focused on enhancing the security of such systems.

\subsection{Related Work} \label{sec:related_work}

Recent work has investigated various vulnerabilities of FL systems. These include backdoor injection \cite{pmlr-v108-bagdasaryan20a, NEURIPS2020_b8ffa41d} and data poisoning \cite{9618642} attacks, both of which can significantly hinder the training process of the FL framework. Additionally, privacy-focused attacks such as the gradient leakage attack \cite{NEURIPS2019_60a6c400, NEURIPS2020_c4ede56b} and model inversion attack \cite{fredrikson2015model} reveal private client information to the attacker. To address these vulnerabilities, various strategies have been proposed, spanning novel aggregation procedures that ensure robustness against poisoned data \cite{9721118}, parameter tuning techniques that enhance model resilience to adversarial examples \cite{Cao_Jia_Gong_2021}, and the adoption of differential privacy methods \cite{9714350, 9069945} to safeguard model integrity and data privacy. More recently, attention has been directed towards the robustness of FL frameworks with different network topologies \cite{raynal2023decentralized}, and has been studied for both semi-centralized and fully decentralized network configurations. However, the impact that adversarial placement has on these vulnerabilities and mitigation strategies remains largely unexplored for the decentralized FL setting.

While the impact of adversarial placement has not been explored in decentralized FL networks, the topic of influence maximization has received significant attention for other network applications. Examples include virus spread in computer networks \cite{4549746, 5340022, singh2018modeling, nwokoye2022epidemic,mirchev2011selective,10.1145/1811039.1811063}, epidemic modeling over contact networks \cite{7029011}, and information (or misinformation) spread in social networks \cite{9679820, wang2023multi, li2023survey}. In each of these scenarios, a common problem is to identify a small subset of network nodes (i.e. spreaders) whose participation in an attack will maximize influence/damage to the network. This problem is in general NP hard, and thus, approximate or heuristic solutions are often considered. In particular, two predominant threads of adversarial selection strategies consistently emerge: \textit{centrality-based attacks} and \textit{distance-based attacks}.

A centrality-based attack involves selecting the spreaders to be the most central nodes in the network. Centrality is a measure of a node's importance within a network, which has been quantified through various metrics, such as degree centrality, betweenness centrality, closeness centrality, and eigenvector centrality  \cite{mirchev2011selective, 10.1145/1811039.1811063, lozano2017optimizing, 7582484}. The rationale for centrality-based attack lies in the assumption that the most central nodes in the network have the best chance of influencing other nodes. In certain network topologies, however, centrality-based attacks may perform poorly. This can occur when the most central nodes are clustered together, and, thus, selecting all of them as spreaders would have significant overlap \cite{Kempe2023}. In such scenarios, distance-based attacks may offer a more effective approach by maximizing the distance between the spreaders \cite{chaoqi2018multi, lin2022graph, binesh2021distance}. Overall, the performance of a given heuristic strongly depends on the topology of the network and the spread model \cite{7582484}.

Unfortunately, many of the above insights developed for network spreading attacks do not directly transfer to attacks on decentralized FL training. Training attacks, like model poisoning, differ from virus/misinformation spreading in that decentralized FL systems have a more complex state space due to factors like the aggregation procedure, heterogeneous datasets, and optimization algorithms. One goal of this paper is understand how well heuristics developed for network spreading problems perform in attacks on decentralized FL systems. Further, we build upon centrality-based attacks and distance-based attacks to construct a hybrid attack for decentralized FL training that demonstrates stronger effects.

\subsection{Summary of Contributions}
\noindent In this paper, we make the following key contributions:
\begin{itemize}
    \item \textbf{Model Poisoning Attack Framework for Decentralized FL}: We initiate the study of adversarial placement for poisoning attacks on decentralized FL systems. We introduce a framework for evaluating how different placement strategies affect decentralized FL training for a given device-to-device topology, learning task, and data distribution (Sec.~\ref{sec:sys_and_atck_model}). 
    
    \item \textbf{Centrality-Based, MaxSpAN-FL, and Hopping-Augmented MaxSpAN-FL Attacks}: We formulate two canonical adversarial placement strategies for decentralized FL. The first is an centrality-based attack, which selects adversarial nodes to maximize an eigenvalue centrality metric. The second is a distance-based attack, coined MaxSpAN-FL (Maximally Spread-out Adversaries in a Network), which (approximately) maximizes the average graph distance between adversaries (Sec.~\ref{sec:new_attack_algorithm}). Finally, we construct a composite attack which augments MaxSpAN-FL with a probabilistic hopping mechanism that takes into the account eigenvector centrality. The resulting algorithm, Hopping-Augmented MaxSpAN-FL, balances between inter-adversarial distance and centrality maximization according to network topology measures (Sec.~\ref{sec:new_attack_algorithm_aug}).
    
    \item \textbf{Experimental Evaluation of Attacks}: We conduct several experiments comparing the potency of these three attacks and a baseline random placement attack (Sec.~\ref{sec:results_and_disc}). Across majority of directed geometric graphs, random graphs, and preferential attachment graphs, we find that Hopping-Augmented MaxSpAN-FL consistently induces the largest degradation on decentralized FL, achieving performance improvements of between $14.5\%$ through $46.2\%$ up to $678.6\%$ performance improvement in some cases. On the other hand, while the centrality-based attack performs well on small graphs, its results for other graph types are more mixed, with it's performance decreasing with with graph size and greatly depending on graph connectivity.

    \item \textbf{Theoretical Justification for Centrality Performance Gap}: We construct a theoretical argument supporting our observation that adversarial placement based on eigenvector centrality is suboptimal (Sec.~\ref{sec:theory}). In particular, we derive a lower bound on the attack effect which is not maximized by choosing the nodes with the largest eigenvalues.
\end{itemize}
This work is an extended version of our conference paper~\cite{piaseczny2024impact}. Compared with the conference version, we (i) extend our MaxSpAN-FL attack methodology to include the hopping component, (ii) add the Preferential Attachment graph type and conduct new experiments involving it, and (iii) provide a theoretical argument as to why centrality-based attacks are suboptimal for decentralized FL.

\section{System and Attack Model}
\label{sec:sys_and_atck_model}
\begin{figure}[t]
\centering
\includegraphics[width=0.48\textwidth]{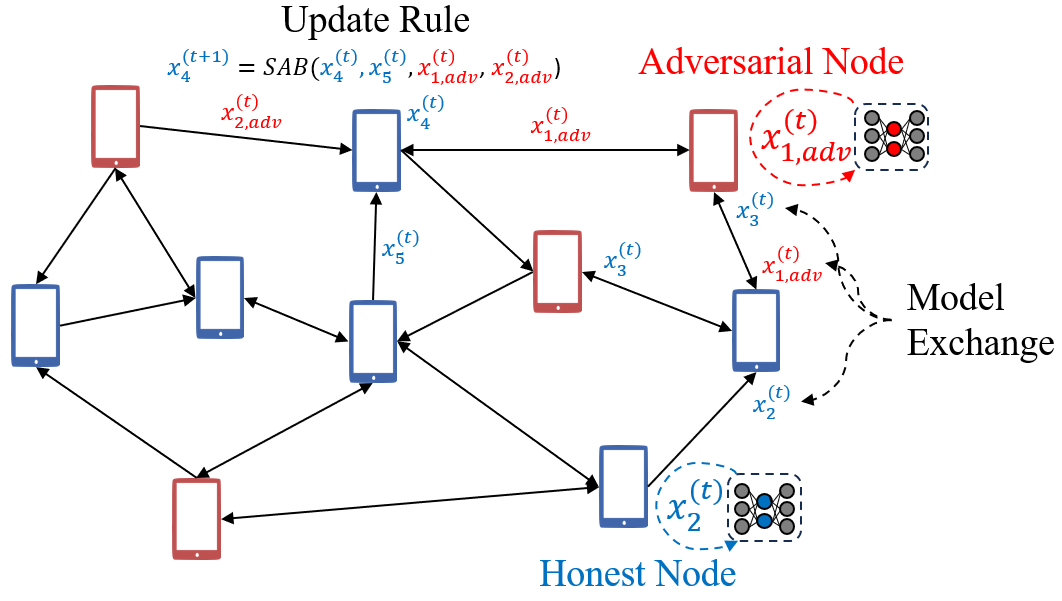}
\caption{Decentralized federated learning with adversarial nodes.}
\label{fig:net_adv}
\vspace{-1em}
\end{figure}

\noindent The decentralized FL system is modeled as follows. As depicted in Fig. \ref{fig:net_adv}, a network of D2D nodes is a strongly connected, time-invariant, directed graph $G = (V, E)$ where $V$ is the set of nodes and $E$ is the set of edges (i.e., communication links) between the nodes. 
The set $V$ can be partitioned into two sets: a set of adversarial nodes $A$ and a set of honest (i.e., non-adversarial) nodes $H:=V \setminus A$. 

The network of D2D nodes participates in a decentralized FL scheme over a time horizon $t=1,2,3,\ldots, N_{\text{epochs}}$. At time $t$, an honest node $i \in H$ trains a local model $x_i^{(t)} \in \mathbb{R}^p$, where $p \geq 1$ is the dimension of the model, using a local data-set $\mathcal{D}_i$ and a local loss function $f_i:\mathbb{R}^p \times \mathbb{R}^{|\mathcal{D}_i|}\rightarrow \mathbb{R}$. 

The goal of the honest nodes is to find a global model $x^*$ that minimizes the global loss function
\begin{equation} \label{eq:global_loss}
   f(x) = \sum_{i \in V\setminus A}f_i(x_i, \mathcal{D}_i).
\end{equation}
The adversaries have a goal opposite of the trusted nodes, namely, to maximize the global loss function (\ref{eq:global_loss}).

We do not assume a central coordinator, and, thus, the honest nodes use distributed averaged consensus for training. Various consensus-based model exchange frameworks \cite{6930814, 4749425, 5936104} have been considered in the literature. We utilize the the S-AB aggregation procedure \cite{9029217} which is a well-known gradient tracking approach. In this setup, at every time $t=1,2,\ldots,N_{\text{epochs}}$, honest node $i \in H$ maintains two parameters: a gradient estimate $y^{(t)}_i$ and a local model $x^{(t)}_i$. These parameters are calculated as follows:
\begin{align}
    x^{(t)}_i = \sum_{j:(j,i)\in E} \frac{x_j^{(t-1)}}{|j:(j,i) \in E| + 1} -\alpha y^{(t-1)}_i, \\
    y^{(t)}_i = \sum_{j:(i,j)\in E} \frac{y_j^{(t-1)}}{|j:(i,j) \in E|} \nonumber \\ + \nabla_{\mathcal{D}_i}f_i(x_i^{(t)}, \mathcal{D}_i) - \nabla_{\mathcal{D}_i}f_i(x_i^{(t-1)}, \mathcal{D}_i)
\end{align}
where $y_i^{(0)}$ and $x_i^{(0)}$ are initial parameters. The above aggregation procedure guarantees that all nodes reach consensus and find an optimal model in the setting where all nodes are honest (i.e., $A$ is empty).

Adversarial nodes do not follow the our established aggregation procedure.
Instead, the adversarial node $i \in A$ mounts an FGSM attack \cite{goodfellow2015explaining} at time $t$ using a local poisoned data-set $\mathcal{D}_{i, adv}^{(t-1)}$ calculated as:\begin{equation}
    \mathcal{D}_{i, adv}^{(t-1)} = \mathcal{D}_i + \epsilon_i \cdot \text{sign}(\nabla_{\mathcal{D}_i}f_i(x_i^{(t-1)}, \mathcal{D}_i))  
\end{equation}\label{eq:fgsm_eq}
where $\epsilon_i > 0$ is the attack power of node $i$. Using the poisoned data, the adversary computes the local model 
\begin{equation} \label{eq:A_model_update}
x^{(t)}_i = x^{(t-1)}_i - \alpha \nabla_{x}f_i(x^{(t-1)}_i, \mathcal{D}_{i, adv}^{(t-1)}).
\end{equation}
Note that the above local model equations state that the adversaries do not coordinate to mount their model poisoning attacks. Lastly, we assume that all adversarial nodes have the same attack power, i.e., $\forall i \in A$, $\epsilon_i = \epsilon$. After training but before time $t+1$, all nodes exchange their local models and global gradient estimations. Adversarial nodes always ignore models from their in-neighbors.


\section{Adversarial Placement Methodology}
\label{sec:methodology}
\noindent In this section, we first introduce a simple baseline algorithm for evaluating adversarial node placement in decentralized FL in Section \ref{sec:baseline_algorithms1}. After this, we present the centrality-based attack and distance-based attack in Section \ref{sec:baseline_algorithms2} and Section \ref{sec:new_attack_algorithm}, respectively.  Finally, we present an attack that borrows ideas from both the centrality-based and distance-based attack in Section \ref{sec:new_attack_algorithm_aug}.

\subsection{Random Placement Algorithm}\label{sec:baseline_algorithms1}
To help evaluate potency of a placement attack, we consider the following random-placement attack as a baseline algorithm for placing adversarial nodes. In a random-placement attack, the adversarial set $A \subset V$ is randomly selected with the size $|A|$ specified in advance. The random-placement attack does not depend on the network topology, making it the simplest form of attack in our study and providing a basic point of comparison for other attacks. 

\subsection{Centrality-based Attack Algorithm}\label{sec:baseline_algorithms2}
The second attack we consider is the network (eigenvector) centrality-based attack. Define $v_i$ as the eigenvector centrality of node $i \in V$, where the centrality vector $\mathbf{v} = (v_1,v_2,\ldots,v_{|V|})$ is computed as $E \mathbf{v} = \lambda_{\mathrm{max}} \mathbf{v}$ where $\lambda_{\mathrm{max}}$ denotes the largest right eigenvalue. In an centrality-based attack, the adversarial set $A$ is formed by selecting the $|A|$ nodes in $V$ with the largest eigenvector centralities. Lastly, we note that eigenvalue centralities can be efficiently computed for large networks using power iteration methods. 




\subsection{MaxSpAN-FL Attack Algorithm}\label{sec:new_attack_algorithm}

The distance-based attack finds an adversarial set $A \subset V$ that maximizes the total distance between $A$ on $G$ $$d_{\text{total}} = \sum_{i \in A} \sum_{\substack{j \in A \\ j>i}}d(i,j)$$ where $d(i,j)$ is the shortest graph hop distance between nodes $i$ and $j$ on $G$. In general, finding a solution to the above problem is computationally infeasible. Thus, we propose an approximate and efficient solution termed the Maximally Spread-out Adversaries in a Network for Federated Learning (MaxSpAN-FL) placement attack. The MaxSpAN-FL attack employs a breadth-first search (BFS) clustering technique to consider the structure of the neighborhood surrounding a node and spread the adversaries throughout the network in an efficient way.

\begin{algorithm}
\caption{MaxSpAN-FL Attack Algorithm}

\begin{algorithmic}[1]
\STATE \textbf{Input:} Graph \( G \), Number of adversaries \( n_{\text{advs}} \)
\STATE \textbf{Output:} List of adversarial nodes \( A \)
\STATE \( \text{Cluster Area } S_{\text{cluster}} \leftarrow \lfloor |G| / n_{\text{advs}} \rfloor \)
\STATE Cluster dictionary \(\mathcal{C}_G \leftarrow \{g:\{\}\}_{g \in G}\) 
\FOR{\(g \in G \)}
    \STATE \( \mathcal{C}_G[g] \leftarrow BFS\_Cluster(G, g, S_{\text{cluster}})\)
\ENDFOR
\STATE Set of adversaries \(A \leftarrow \{\}\)
\STATE Set of honest nodes \(H \leftarrow G\)
\WHILE{\( |A| < n_{\text{advs}} \)}
    \STATE \( o_{\min} \leftarrow \infty \)
    \STATE \( a_{\text{best}} \leftarrow \text{None} \)
    \FOR{\(g \in H \)}
        \STATE \(o = \left|\mathcal{C}_G[g] \cap \left(\underset{a \in A}{\cup}\mathcal{C}_G[a]\right)\right|\)
        \IF{\(o < o_{\min}\)}
            \STATE \( o_{\min} \leftarrow o\)
            \STATE \( a_{\text{best}}\leftarrow g \)
        \ENDIF
    \ENDFOR
    \STATE \( A \leftarrow A \cup {a_{\text{best}}}\)
    \STATE \( H \leftarrow H \setminus {a_{\text{best}}}\)
\ENDWHILE
\RETURN \( A \)
\end{algorithmic}

\end{algorithm}\label{alg_MaxSpAN-FL}

As outlined in \ref{alg_MaxSpAN-FL}, the MaxSpAN-FL algorithm initially calculates influence regions for all nodes \( g \in G \) using BFS. BFS is run until the number of nodes in a given cluster exceeds the pre-determined cluster area \( S_{\text{cluster}} \). This area is calculated by dividing the total number of nodes \( n_{\text{clients}} \) by the number of adversaries \( n_{\text{advs}} \), ensuring that each adversary has a roughly equal area of influence.

Once these influence regions are established, the algorithm enters an iterative phase where it selects adversarial nodes \( a_{\text{best}} \) in a way that minimizes the overlap between their influence regions. Specifically, for each node $g$ still considered `honest', the algorithm calculates the overlap \( o \) of its influence region with those of the already selected adversarial nodes. This overlap is quantified as 
\begin{equation}
    o = \left|\mathcal{C}_G[g] \cap \left( \bigcup_{a \in A} \mathcal{C}_G[a] \right) \right|.
\end{equation}
\rv{where $\mathcal{C}_G[g]$ is the cluster created around node $g$ using BFS.}
The node with the least overlap is then selected as the next adversarial node, added to the set \( A \) of adversarial nodes, and removed from the set \( H \) of honest nodes.

We remark that the algorithm introduces a degree of randomness in the placement of adversarial nodes. The first node is chosen completely at random, providing a starting point for the algorithm. Subsequent nodes are selected based on minimizing overlap.


\subsection{Hopping-Augmented MaxSpAN-FL}\label{sec:new_attack_algorithm_aug}
Building upon the MaxSpAN-FL placement attack, we propose a new modification deemed Hopping-Augmented MaxSpAN-FL that incorporates adaptive hopping techniques in order to optimize attack performance for differing network topologies. The augmented algorithm is outlined in \ref{alg_MaxSpAN-FL_w_hop}.

\begin{algorithm}
\caption{Hopping-Augmented MaxSpAN-FL}

\begin{algorithmic}[1]
\STATE \textbf{Input:} Graph \( G \), Number of adversaries \( n_{\text{advs}} \), Decision Boundary Parameters \( \boldsymbol{\alpha} \), Decay Parameter \(\lambda\)
\STATE \textbf{Output:} List of adversarial nodes \( A \)
\STATE Set of adversaries \(A \leftarrow \{\}\)
\STATE \(A \leftarrow \text{MaxSpAN-FL}( G,  n_{\text{advs}} )\)
\STATE \(\sigma^2_{\text{deg}} \leftarrow \text{degree variance}(G)\)
\STATE \(d_{\min} \leftarrow \min_{g \in G} \text{degree}(g)\)
\STATE \(d_{\max} \leftarrow \max_{g \in G} \text{degree}(g)\)
\STATE \(\hat{\sigma}^2_{\text{deg}} \leftarrow (\sigma^2_{\text{deg}} - d_{\min}) / (d_{\max} - d_{\min})\)
\FOR{\(a \in A\)}
\STATE \(t \leftarrow 0\)
\STATE \(c_a \leftarrow \text{clustering coefficient}(G, a)\)
\STATE \(c_{\min} \leftarrow \min_{g \in G} \text{clustering coefficient}(G, g)\)
\STATE \(c_{\max} \leftarrow \max_{g \in G} \text{clustering coefficient}(G, g)\)
\STATE \(\hat{c}_a \leftarrow (c_a - c_{\min}) / (c_{\max} - c_{\min})\)
\REPEAT
    \STATE \(p \leftarrow 1 / (1 + \exp(\alpha_0 \cdot (\hat{c}_a + \alpha_1 ) \cdot (\hat{\sigma}^2_{\text{deg}}  + \alpha_2))  \cdot \exp(-\lambda \cdot t / \log(|G|))\)
    \STATE \(N \leftarrow \text{neighbors}(a) \setminus A\)
    \STATE \(a \leftarrow \argmax_{n \in N} {\text{centrality}(n)}\)
    \STATE \(t \leftarrow t + 1\)

\UNTIL{\(\text{random}() \geq p\)} 
\ENDFOR\RETURN \( A \)
\end{algorithmic}

\end{algorithm}\label{alg_MaxSpAN-FL_w_hop}

The key change of the Hopping-Augmented MaxSpAN-FL algorithms lies in the incorporation of an adaptive hopping mechanism. This mechanism is strategically placed on top of the original BFS-based adversarial selection process established by MaxSpAN-FL. After the initial placement of adversarial nodes using the MaxSpAN-FL algorithm, the hopping mechanism fine-tunes their positions within the network. A "hop" in this context defines a calculated movement of an adversarial node from its initial position to a neighboring node within the graph. The algorithm dynamically determines the number of hops based on the normalized measure of node degree variation \(\hat{\sigma}^2_{\text{deg}}\) and the normalized clustering coefficients of nodes within the network \(\hat{c}_a\). Along with decision boundary parameters \(\boldsymbol{\alpha} = \begin{bmatrix} \alpha_0 & \alpha_1 & \alpha_2 \end{bmatrix}^T\), and a decay parameter $\lambda$, the probability of a hop happening for an adversarial node $a \in A \subseteq G$ can be described as:
\begin{equation}
    P_{\text{hop}}(a) = \frac{e^{-\frac{\lambda}{|G|}}}{1 + e^{\alpha_0 (\hat{c}_a + \alpha_1)(\hat{\sigma}^2_{\text{deg}} + \alpha_2)}}
\end{equation}
This adaptive measure is essential to accommodate different structural characteristics of various network types. Networks with large clustering coefficients and node degree variation will have fewer hops and the adversarial placement will rely more heavily on distance-maximization. In contrast, networks with small clustering coefficients and node degree variation will have more for hops per adversary.

\begin{figure}[t]
\centering
\includegraphics[width=0.48\textwidth]{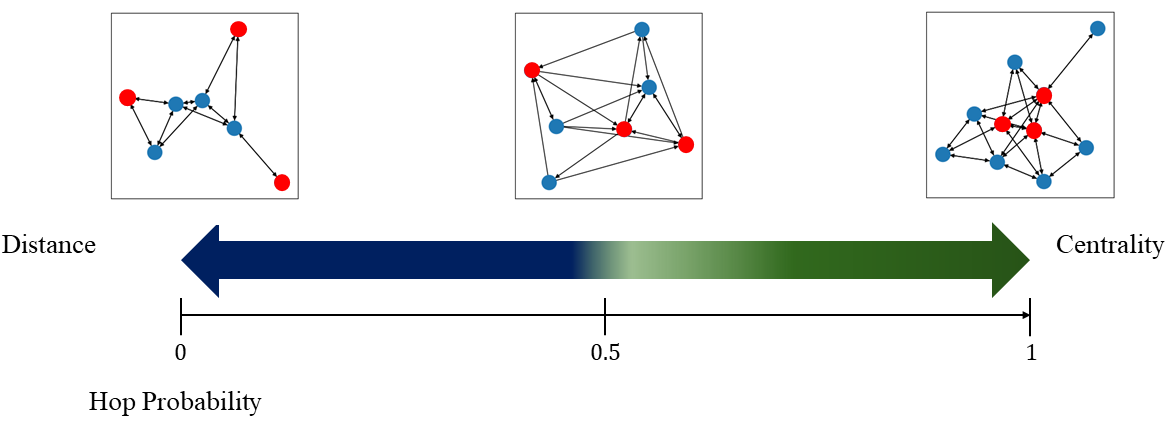}
\caption{Trade-off between distancde-based and centrality-based node selection depending on the hop probability. As the hop probability increases, the adversaries will tend to occupy more central nodes.}
\label{fig:dist_cent_tradeoff}
\vspace{-1em}
\end{figure}

Once a hop happens, the algorithm uses a centrality-based selection of neighboring nodes for deciding new adversarial positions. The new adversarial node will always be a neighbor of the previously chosen adversary with the highest eigenvector centrality, which can be described as
\begin{equation}
    a_{i + 1} \leftarrow \argmax_{n \in \text{neighbors}(a_i) \ A} \text{centrality}(n)
\end{equation}
By utilizing stochastic hopping mechanism to central nodes, the extended algorithm creates an effective trade-off between distance-based and centrality-based node selection, which is illustrated in Figure \ref{fig:dist_cent_tradeoff}.

\section{Results and Discussion}
\label{sec:results_and_disc}
\noindent Here, we first discuss the experimental setup in Section~\ref{exp_setup}, which includes the network architectures, model architectures, and datasets used for evaluation. Then, in Section~\ref{exp_results}, we present the experiments conducted and discuss the corresponding results for various hyperparameter settings.
\subsection{Experimental Setup}\label{exp_setup}
In our experiments, we consider the decentralized FL model of Section \ref{sec:sys_and_atck_model}. We evaluate this model for distinct network topologies in which the graph $G$ is drawn at random. We consider the following random graph distributions: 
\begin{itemize}
    \item Erdős-Rényi (ER) graphs
    \item Directed Geometric (DG) graphs 
    \item Preferential Attachment (PA) graphs
\end{itemize} To assess the effectiveness of the proposed MaxSpAN-FL and Hopping-Augmented MaxSpAN-FL Attacks we synthesize networks with varying hyperparameters, such as network connectivity or network size. For each graph distribution, we generate 20 graph realizations.

The network hyperparameters we consider are as follows:
\begin{itemize}
    \item Connectivity parameters for 25-node networks for both graph distributions. For the DG graphs, we randomly assign nodes locations in a 2D unit square and vary the connection radius parameter $r$ to $0.2$, $0.4$, and $0.6$ to simulate different levels of network density. For the ER graphs, we adjust the edge creation probability $p$, setting it to $0.1$, $0.3$, and $0.5$. For the PA graphs, we adjust the initial graph size that new nodes are attached to. We consider initial graph sizes of $1$, $2$, and $6$ for differing network densitites. 
    \item Size parameters for the DG graph with $r = 0.2$ and the PA graph with initial graph size of $1$. We consider network sizes of 10, 25, 50, and 100, and for each network we consider different adversarial percentages: $10\%$ and $20\%$.
\end{itemize}

Following network generation, each node trains an image classification model. We use the Fashion-MNIST (FMNIST) dataset for evaluation, distributing it across nodes in both IID and Non-IID manners, with the latter having 3 classes per node. Tailored Convolutional Neural Network (CNN) model architecture serves as our classifier. The model architecture consists of two convolutional layers, each succeeded by a dropout layer, a ReLU activation function, and a max-pooling operation, culminating in two dense layers for classification. 

Each node executes a predefined number of local training iterations on their respective model. Upon completion, the models are distributed amongst neighboring nodes via the consensus aggregation algorithm described in Section \ref{sec:sys_and_atck_model}. We use Adam optimizer for training.

In all cases, we evaluate the random baseline attack, the Eigenvector-centrality based attack, as well as the MaxSpAN-FL and Hopping-Augmented MaxSpAN-FL attacks for selecting the group of adversarial nodes that run FGSM as described in \ref{sec:sys_and_atck_model}. \footnote{The source code supporting our experiments is publicly available at \url{https://github.com/Adampi210/MaxSpANFL_atck_code_data.git}.}

\subsection{Experimental Results}\label{exp_results}
\subsubsection{Effects of Graph Distribution}
\begin{figure*}[t]
  \centering
\includegraphics[scale=0.48]{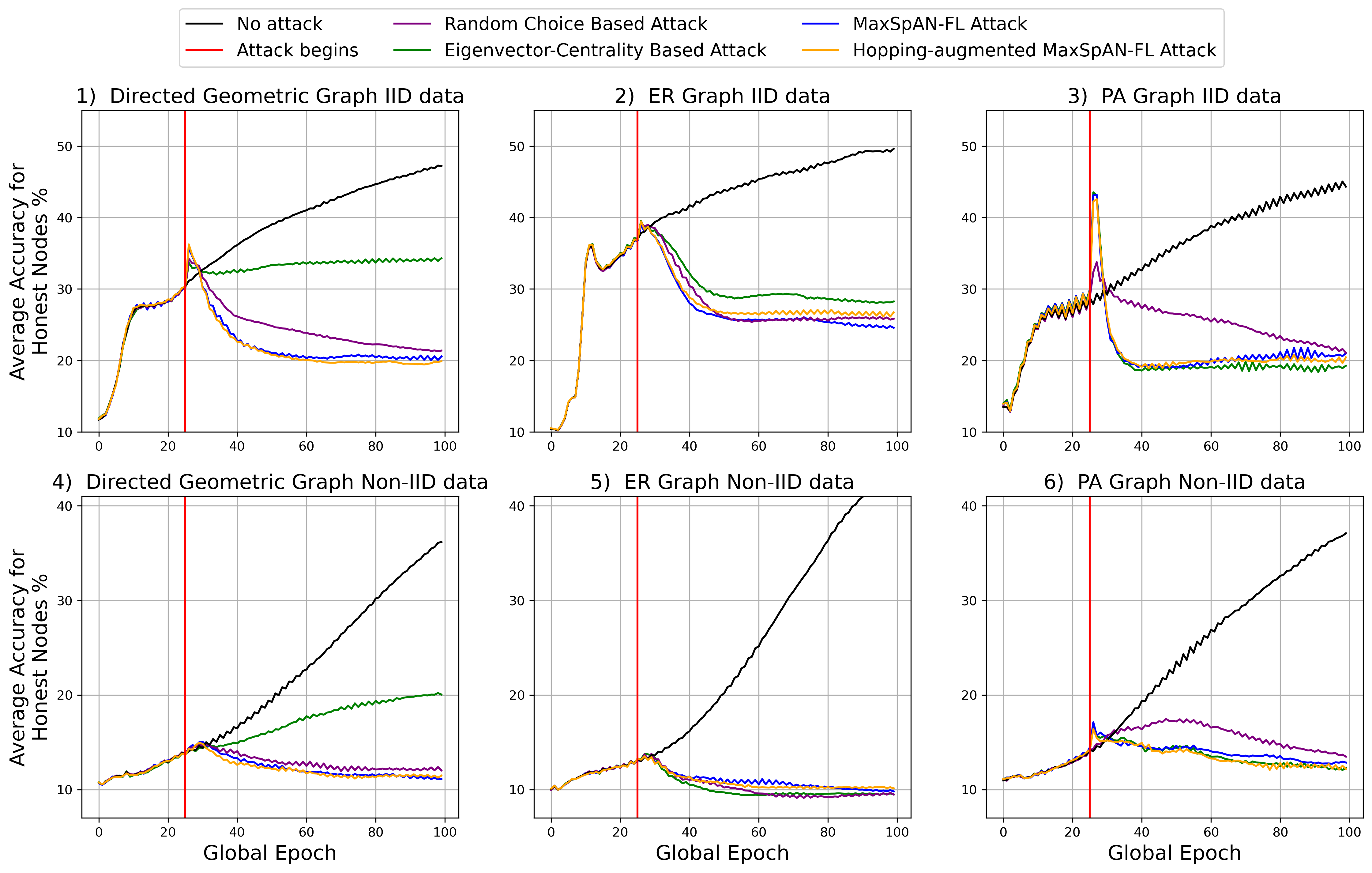}
  \caption{Average testing accuracy of honest nodes in 25-node networks, comparing Directed Geometric graphs with connection radius \( r = 0.2 \), Erdős–Rényi graphs with edge probability \( p = 0.3 \), and Prefrential Attachment graphs with initial graph size of 1 node, for both IID and Non-IID data distributions. The effects of various attack placement strategies on the network's performance are illustrated. Adversarial percentage is $20\%$. Lower accuracy is better.}
  \label{fig:main_result}
  \vspace{-1.3em}
\end{figure*}

Figure \ref{fig:main_result} illustrates that the potency of adversarial placement attacks depends on both the network's graph distribution and the data distribution among clients.

In directed geometric graphs with IID data, the MaxSpAN-FL and the Hopping-Augmented MaxSpAN-FL attacks outperformed alternative strategies, achieving final test accuracies of \(20.6\%\) and \(19.9\%\), respectively. These results surpassed the \(21.4\%\) and \(34.3\%\) final accuracies achieved by random node and Eigenvector-centrality based attacks, respectively. Both attacks reached their lowest accuracy around the 60th global epoch, which is not achieved by other methods even by the 100th epoch, highlighting the significance of strategic adversarial placement in directed geometric graphs.

The performance in the ER graph case provides further insights. The similar effectiveness observed across placement attack strategies can be attributed to the degree uniformity in ER graphs, where all nodes tend to have similar centrality scores and any two nodes are likely to be separated by only a short path. Consequently, the random placement attack performs comparably to both MaxSpAN-FL attacks. The underperformance of the Eigenvector-centrality based attack likely stems from variance in the averaging of results, with the performance differences being significantly less pronounced than in the directed geometric graph scenario. This trend is observed for both the IID and Non-IID data distributions.

In the case of preferential attachment graphs, final performance metrics were similar across all attacks, with the Eigenvector-centrality based attack achieving the lowest final accuracy of \(19.3\%\). However, the dynamics of the attacks varied. The Random-choice attack, for instance, achieved its final accuracy of \(21.1\%\) at a much slower pace and later epoch compared to other strategies, indicating the existence of an optimal attack choice for this scenario. Notably, although the Eigenvector-centrality attack was the most effective overall, the Hopping-Augmented MaxSpAN-FL was a close second, achieving \(20.4\%\) final accuracy and slightly outperforming the purely distance-focused attack, reflecting its hybrid approach based on interaction between distance and centrality.

Overall, the MaxSpAN-FL and the Hopping-Augmented MaxSpAN-FL consistently achieved the best or comparable performances across all scenarios and exhibited the fastest or near-fastest rate of decreasing test accuracy. Interestingly, the experiments also demonstrated that the data distribution among clients significantly influences the performance gap between placement attack strategies. In Non-IID scenarios, this gap notably narrows, potentially due to the complexities of class distribution in adversarial node placement and varying degrees of model convergence at the time of attack. Notably, test accuracies of attacked clients in Non-IID settings are substantially lower than in IID scenarios, further underscoring the role of data distribution and model optimality in shaping attack outcomes.

\subsubsection{Effects of Connectivity} \label{sec:eff_conn}
\begin{figure}[t]
\centering

\begin{subfigure}{\columnwidth}
  \centering
  \includegraphics[width=\linewidth]{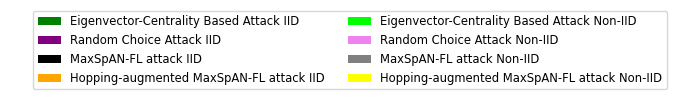}
\end{subfigure}%
\vspace{-0.15em}
\begin{subfigure}{\columnwidth}
  \centering
  \includegraphics[width=\linewidth]{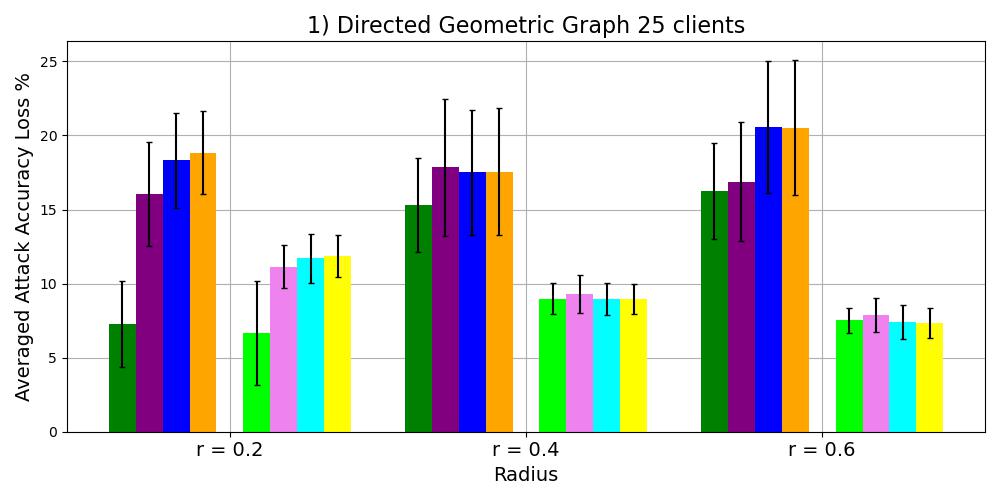}
\end{subfigure}
\vspace{-0.15em}
\begin{subfigure}{\columnwidth}
  \centering
  \includegraphics[width=\linewidth]{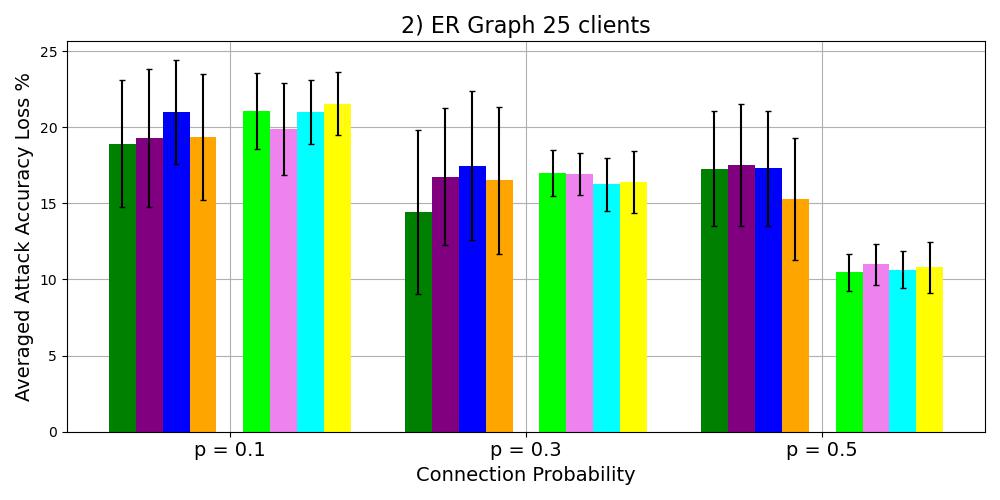}
  
\end{subfigure}
\vspace{-0.15em}
\begin{subfigure}{\columnwidth}
  \centering
  \includegraphics[width=\linewidth]{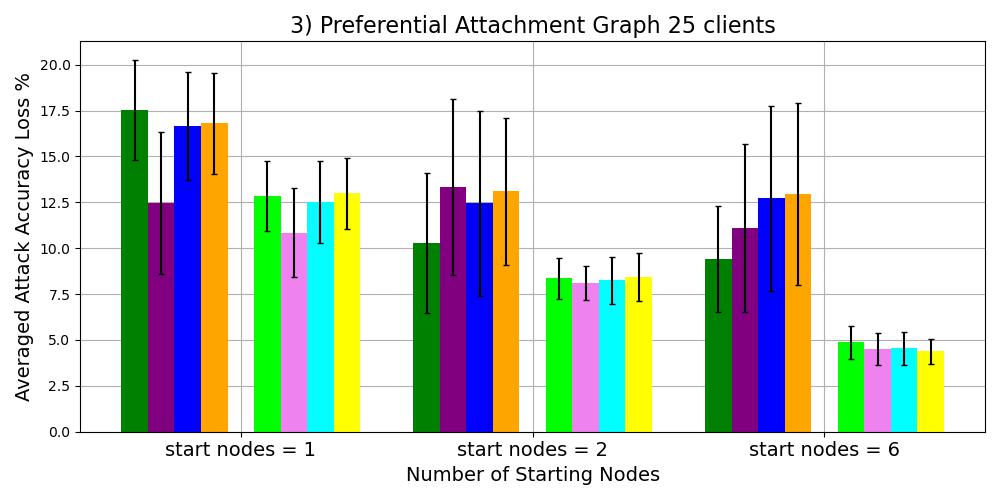}
  
\end{subfigure}
\vspace{-1.5em}
\caption{Average Attack Accuracy Loss (AAL) for Directed Geometric, ER, and Preferential Attachment graphs with 25 nodes and $20\%$ adversaries for various connectivity parameters. Higher AAL corresponds to a more effective attack.}
\label{fig:connectivity_results}
\vspace{-1.5em}
\end{figure}

\rv{Network connectivity has an important effect on the information exchange between clients in a decentralized FL systems.}\footnote{\rv{Although we focus on network-level properties like connectivity, our framework can easily be extended to model properties of the wireless physical layer (e.g., bandwidth, transmission power) as they relate to connectivity. For further discussion, see \cite{Yao2024}.}} Figure \ref{fig:connectivity_results} demonstrates the influence of network connectivity on the potency of different placement attacks, considering both IID and Non-IID data distributions in networks with 25 nodes. The analysis of this setting is based on the Attack Accuracy Loss ($\text{AAL}$) measure $$\text{AAL} = \overset{N_{\text{epochs}}}{\underset{i = t_{\text{attack}}}{\sum}}\left[\text{(\% acc no attack)[i]} - \text{(\% acc with attack)[i]}\right]$$
which measures the impact an attack has on test accuracy over all epochs. Higher AAL indicates better attack potency.
Attack performance advantage is computed as: $$\frac{\text{AAL}_{\text{best attack}}-\text{AAL}_{\text{next best attack}}}{\text{AAL}_{\text{next best attack}}} \times 100\%$$

In Directed Geometric graphs with varying connectivity levels (radius values \( r \in \{0.2, 0.4, 0.6\} \)), both MaxSpAN-FL and the Hopping-Augmented MaxSpAN-FL attacks exhibit superior performance, particularly in the \( r = 0.2 \) and \( r = 0.6 \) scenarios under IID conditions, outperforming the Random choice attack by \( 17.2\% \) and \( 21.8\% \) respectively. The performance gap narrows in the \( r = 0.4 \) scenario, with both attacks slightly trailing the Random choice attack by approximately \( 1.8\% \). These findings underscore the resilience of the MaxSpAN-FL and Hopping-Augmented MaxSpAN-FL methods across different levels of connectivity in DG graphs. Of note is the fact that the Eigenvector-centrality based attack severely underperforms other attack schemes for all connectivity settings in DG graphs. 

For ER graphs, we study AAL across various connection probabilities \( p \in \{0.1, 0.3, 0.5\} \). The MaxSpAN-FL attack displays its greatest advantage in the \( p = 0.1 \) case, surpassing the next best attack by approximately \( 9\% \). The Hopping-Augmented MaxSpAN-FL attack, however, shows slightly lower AAL across all IID scenarios. Overall, this result suggests that top attack schemes perform similarly for this graph distribution, especially for high graph density.

For Preferential Attachment graphs, we study AAL over various number of starting nodes. In sparse conditions, the most effective attack is the Eigenvector-centrality based attack, closely followed by both the Hopping-Augmented and vanilla MaxSpAN-FL attacks, which perform about \(4.3\%\) worse. As connectivity increases, the performance of the centrality-based attack declines sharply. Remarkably, in medium-dense conditions (starting nodes number equal to $2$), the random attack outperforms all others, indicating that other node selection strategies, potentially involving different graph and node properties, may be optimal. In very dense conditions, however, both Hopping-Augmented MaxSpAN-FL and MaxSpAN-FL attacks are most effective, significantly outperforming the Random choice attack by \(16.7\%\) and \(14.5\%\) respectively.

These results not only affirm the performance edge and competitiveness of the Hopping-Augmented and vanilla MaxSpAN-FL attacks across diverse network configurations but also highlight several insights. Notably, in Non-IID scenarios for all graph distributions, the peak AAL consistently decreases with increased connectivity—a trend less pronounced in IID settings. This is likely due to enhanced information exchange among honest clients in denser networks. In highly connected ER graphs, the performance differences between top attacks diminish, suggesting that node placement becomes less critical. Additionally, the Eigenvector-centrality based attack improves its performance by $123.4\%$ in DG graphs under IID conditions as connectivity increases, but sees a decrease of $46.2\%$, highlighting the fact that the impact of connectivity can vary significantly depending on the graph distribution.

\subsubsection{Effects of Network Size and Number of Adversaries}
\begin{figure}[t]
\centering
\begin{subfigure}{\columnwidth}
  \centering
  \includegraphics[width=\linewidth]{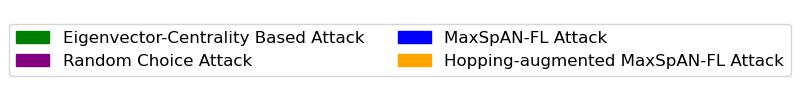}
\end{subfigure}%
\vspace{-0.15em}
\begin{subfigure}{\columnwidth}
  \centering
  \includegraphics[width=\linewidth]{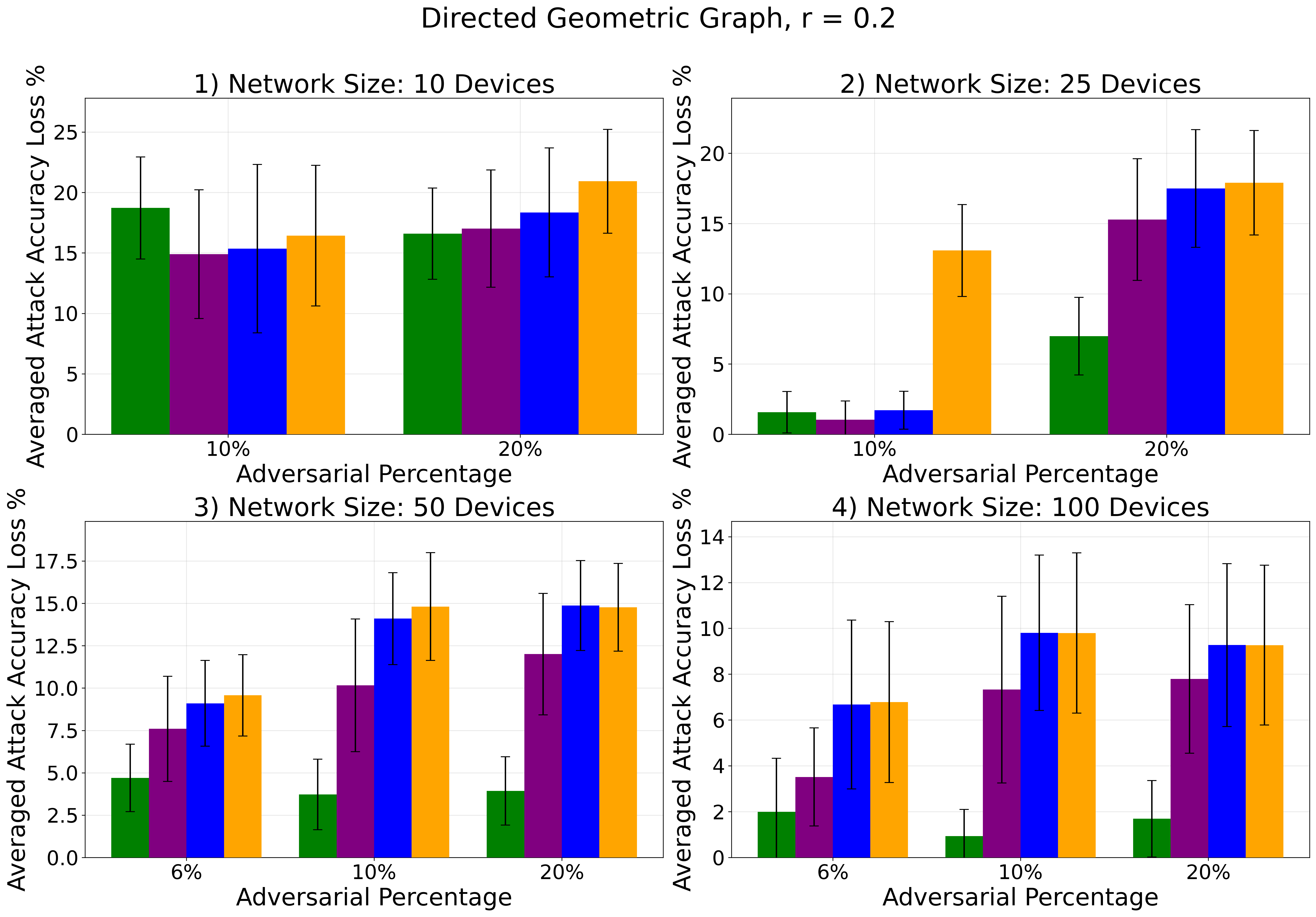}
\end{subfigure}%
\vspace{0.4em}
\begin{subfigure}{\columnwidth}
  \centering
  \includegraphics[width=\linewidth]{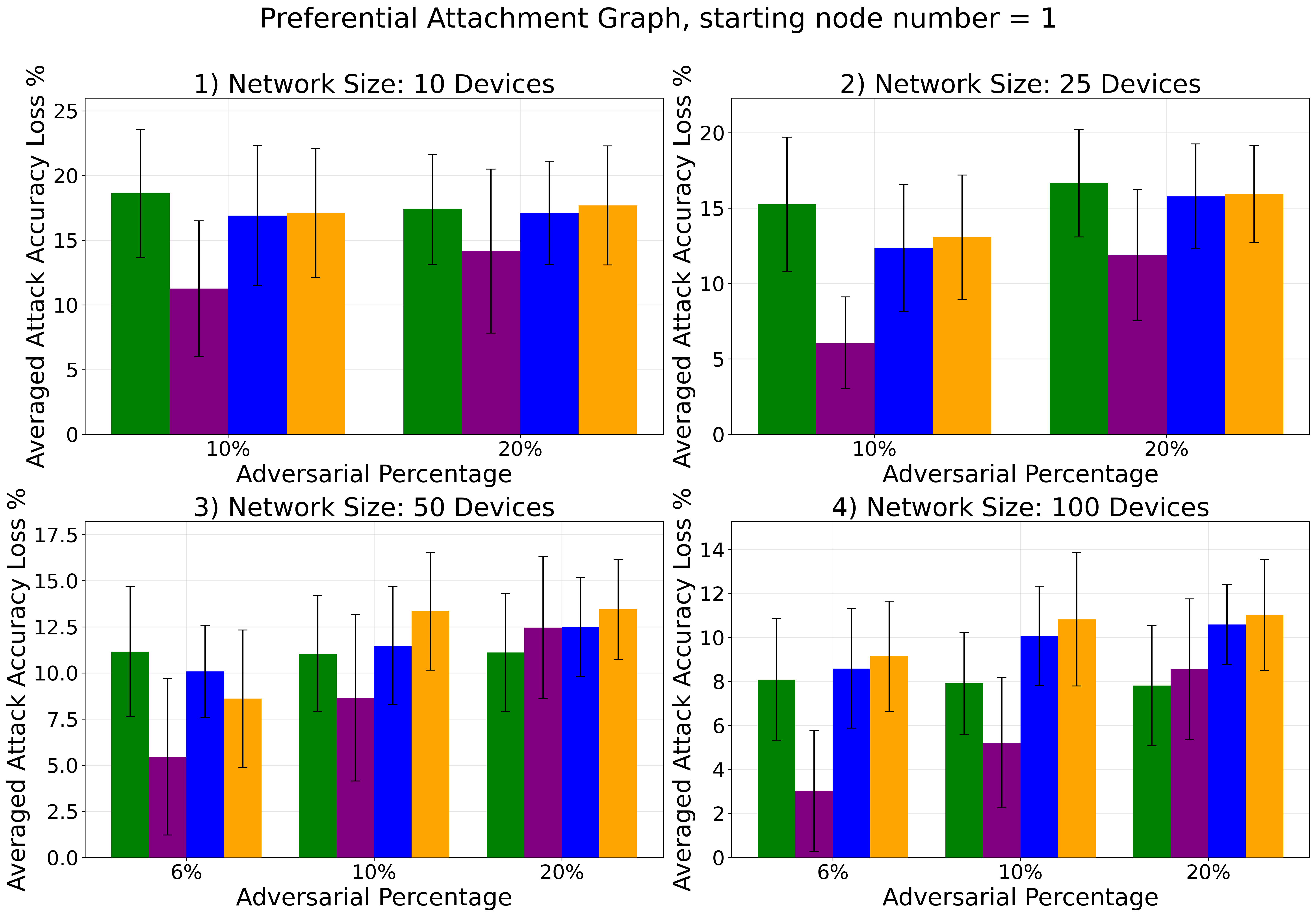}
\end{subfigure}%
\vspace{-0.5em}
\caption{Average Attack Accuracy Loss (AAL) for Directed Geometric graphs with $r = 0.2$ and Preferential Attachment graphs with starting node number $= 1$ and IID data distribution, for different network sizes and number of adversaries. Higher AAL corresponds to a more effective attack.}
\label{fig:network_sizes}
\vspace{-2em}
\end{figure}

The impact of network size and the number of adversarial nodes on AAL is illustrated in Figure \ref{fig:network_sizes}. 

We focus on Directed Geometric graphs with a connectivity radius \( r = 0.2 \) and Preferential Attachment graphs with the number of starting nodes set to 1.\footnote{We exclude ER graphs, as we have observed that AAL is similar across all attack strategies for ER graphs (e.g., see Figure \ref{fig:connectivity_results}.2). } We focus on the setting where data is IID. 

In the DG graphs, the Hopping-Augmented MaxSpAN-FL attack surpasses other methods in nearly all examined cases. For larger networks, its performance aligns closely with the vanilla MaxSpAN-FL, providing considerable improvements over the Random choice attack, with a maximum improvement of around \( 45.3\% \) in the 50-node network with 5 adversaries. However, its distinct advantages are more pronounced in smaller networks with fewer adversaries, where moderate hopping distances allow it to target locally central nodes effectively, outperforming the next best attack by approximately \( 678.6\% \) in the 25-node network with 2 adversaries. Generally, the Hopping-Augmented MaxSpAN-FL achieves the best performance across these graph distributions, except in a 10-client network with a single adversary where the Eigenvector-centrality based attack excels, outperforming the Hopping-Augmented MaxSpAN-FL by \( 13.1\% \). 
Despite this, the augmented algorithm still outperforms both its vanilla version and the Random choice attack, and in general demonstrates an effective attack across different graph sizes. Its performance benefits underscore not only the effectiveness of the proposed placement strategy, but also significant advantages of the hopping mechanism for certain graph configurations.  

In the PA graph scenario, the Hopping-Augmented MaxSpAN-FL attack \rv{mostly} outperforms the regular MaxSpAN-FL, offering up to a \(16.0\%\) improvement in the 50-node network with 5 adversaries. While the Eigenvector-centrality based attack outperforms the Hopping-Augmented MaxSpAN-FL in networks with 10 and 25 nodes, its effectiveness significantly decreases \rv{ with larger numbers of adversarial nodes}, allowing our proposed algorithm to dominate. Indeed, for increasing \rv{number of adversaries in larger graphs}, both the Hopping-Augmented and regular MaxSpAN-FL attacks achieve better or equally good results compared to other attack methods.

Both cases reveal intriguing trends, such as the minimal performance benefit of increasing the adversarial percentage in larger graphs. Moreover, in both scenarios, centrality-based approaches perform well in smaller networks \rv{and with fewer total adversaries}.

Also, of note is the fact that while the centrality-based approach loses effectiveness for larger \rv{adversarial numbers}, the Hopping-Augmented MaxSpAN-FL does not suffer from this problem, which can be explained by a combination of two factors: (1) In some cases, the network properties prevent any hopping, allowing the augmented attack to match the vanilla attack's performance. (2) In other cases, maximizing the localized centrality of the chosen clusters appears to be better than maximizing the global node centrality. 

\rv{

\subsubsection{Centrality Measures Comparison} \label{sec:cmc} In the previous experiments we utilized graph properties and structure to conduct the experiments. The proposed MaxSpAN-FL, Hopping-Augmented MaxSpAN-FL, and Eigenvector-centrality based attacks utilize sophisticated algorithms to select nodes for attack. These proposed algorithms take as input a variety of graph-related information when selecting the adversarial nodes, which leads to their high complexity. This then poses a question of whether similar favorable results could be achieved using simpler methods that utilize basic graph properties. To answer that question we run experiments utilizing a modified version of the centrality based attack, using the degree centrality instead of eigenvector centrality to select adversaries. The degree centrality is calculated as a sum of outgoing and incoming edges for a given node. Using this scheme, top-k nodes with highest degrees are chosen as the adversaries. The result can be seen in Figure \ref{fig:deg_centrality}.

\begin{figure}[t]
\centering
\begin{subfigure}{\columnwidth}
  \centering
  \includegraphics[width=\linewidth]{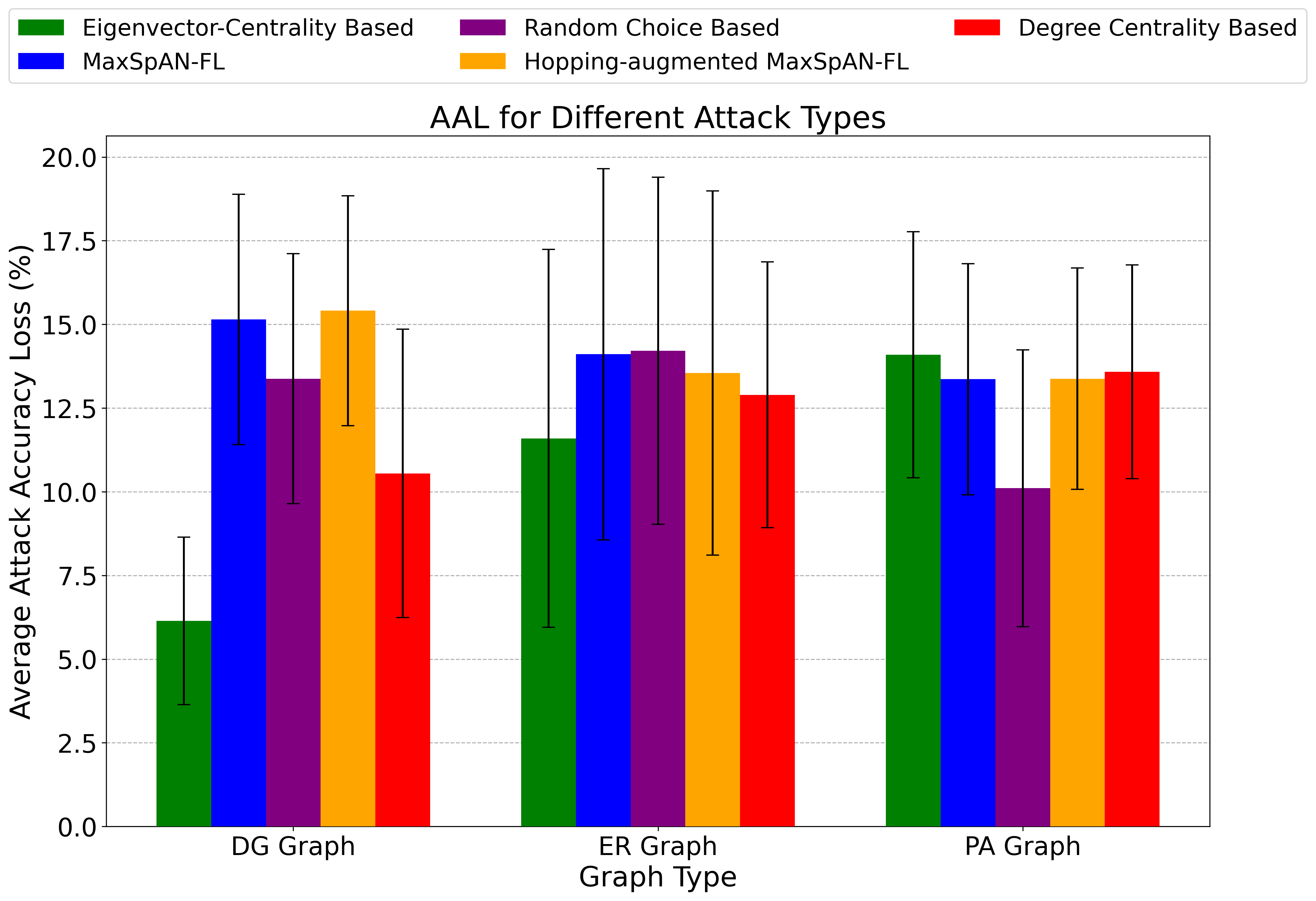}
\end{subfigure}%
\vspace{-0.5em}
\caption{\rv{Average Attack Accuracy Loss (AAL) for Directed Geometric graphs with $r = 0.2$, ER graphs with $p = 0.3$ and Preferential Attachment graphs with starting node number $= 1$ and IID data distribution, for different attack types including the degree-centrality-based attack. Higher AAL corresponds to a more effective attack.}}
\label{fig:deg_centrality}
\vspace{-2.5em}
\end{figure}

The results demonstrate that even with the introduction of a simpler selection scheme, the proposed MaxSpAN-FL and Hopping-Augmented MaxSpAN-FL outperform or match the best performing attack for all the 3 tested graph types. In the DG graph as expected both schemes outperformed all other attacks by considerable margins, and in fact the degree centrality based attack performed worse than the random selection attack, similarily to the Eigenvector-centrality based attack. In the ER graph case, all attack performed similarly well, as expected. Finally, in the PA graph, the degree centrality attack performed best out of all the 3 cases, slightly better than the Hopping-Augmented and regular MaxSpAN-FL but a little worse than the Eigenvector-centrality based attack. Despite this good performance in the PA case, the degree centrality attack does not achieve as good general results on a variety of graphs as the MaxSpAN-FL based schemes, and is not as easy to analyze as the Eigenvector-centrality based attack. This makes the degree based attack an inferior option for deployment and results in the attack having limited applicability in research. While this attack does not account for all simplified selection schemes, it demonstrates that simple utilization of rudimentary graph properties might be insufficient to create an optimal adversarial not selection algorithm. 

\subsubsection{Effects of Link and Node Failure} \label{sec:ELNF} We further conduct experiments that are designed to closer resemble real-world scenarios. We extend our analysis to consider a dynamically changing network topology scenario. While we don't incorporate specific physical properties such as in \cite{Yao2024}, we generalize our dynamic networks model to account for link and node failures. In the experiments, after the adversary selects the attack nodes using one of the proposed algorithms, nodes and links will fail according to set failure probabilities. After the failed nodes and edges (links) are removed from the network the rest of the learning process proceeds normally with the new graph. We evaluate 4 settings, each with different probabilities of node and link failures, summarized in Table \ref{tab:link_fail_param}
\begin{table}[htbp]
\centering
\caption{\rv{Network Dynamics Settings: Node and Link Failure Probabilities}}
\label{tab:link_fail_param2}
\begin{tabular}{|c|c|c|}
\hline
\multirow{2}{*}{Setting Name} & Node Failure & Link (Edge) Failure \\
 & Probability & Probability \\
\hline
Low Dynamics & 0.1 & 0.02 \\
\hline
Mild Dynamics & 0.15 & 0.05 \\
\hline
Moderate Dynamics & 0.20 & 0.1 \\
\hline
High Dynamics & 0.3 & 0.2 \\
\hline
\end{tabular}
\end{table}\label{tab:link_fail_param}

The results of those experiments for the Directed Geometric and Preferential Attachment graphs \footnote{We exclude ER graphs as all attacks performed similarly in that scenario.} are demonstrated in Figure \ref{fig:dynamic_graphs}.

\begin{figure}[t]
\centering
\vspace{-0.6em}
\begin{subfigure}{\columnwidth}
  \centering
  \includegraphics[width=\linewidth]{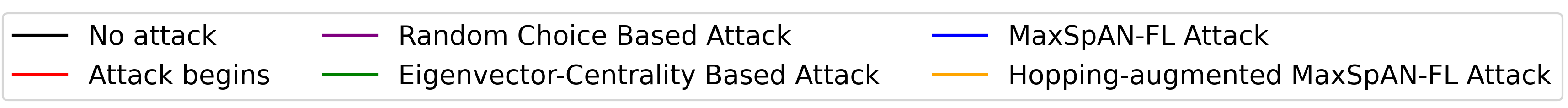}
\end{subfigure}%
\vspace{0.01em}
\begin{subfigure}{\columnwidth}
  \centering
  \includegraphics[width=\linewidth]{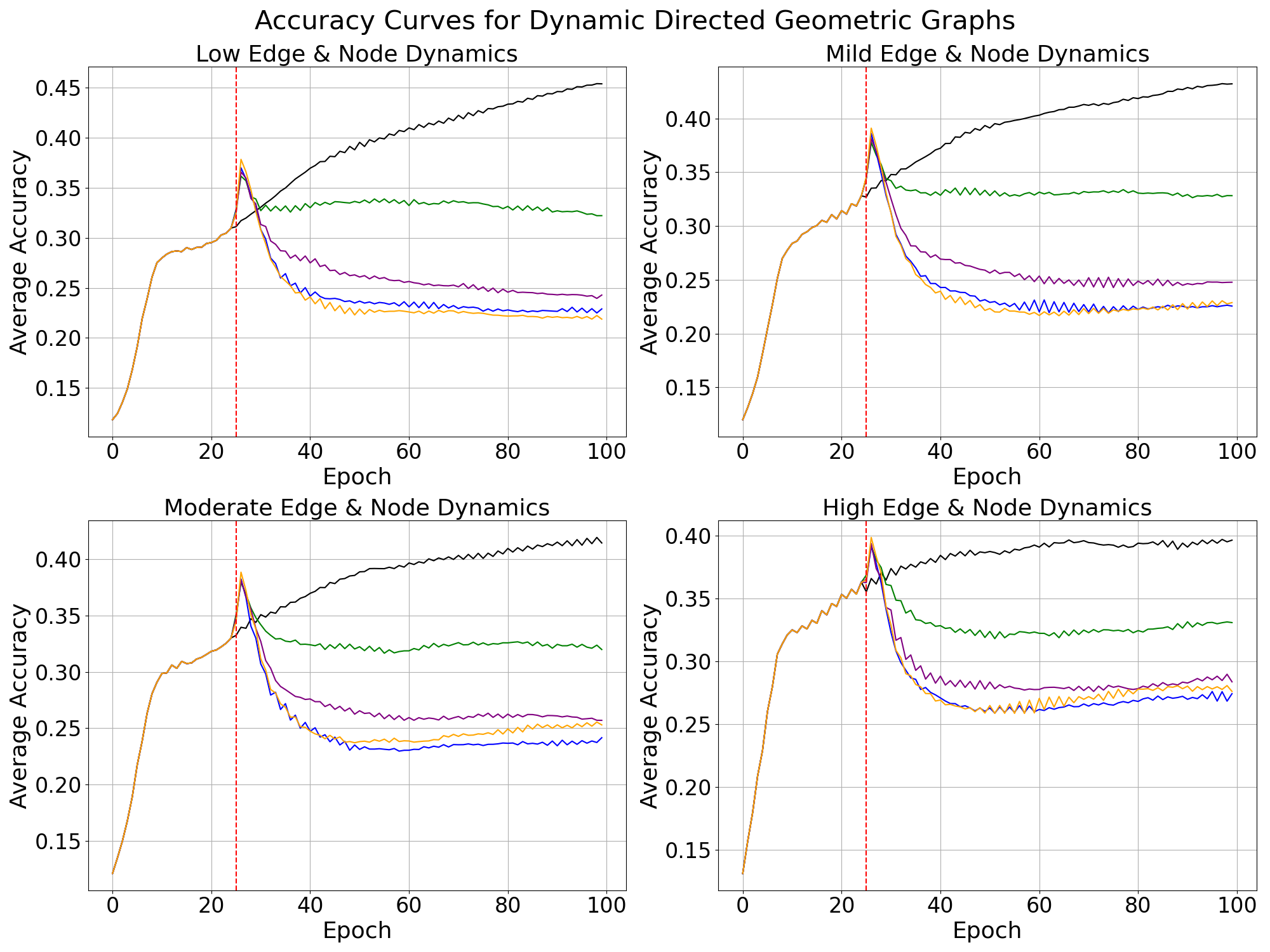}
\end{subfigure}%
\vspace{0.01em}
\begin{subfigure}{\columnwidth}
  \centering
  \includegraphics[width=\linewidth]{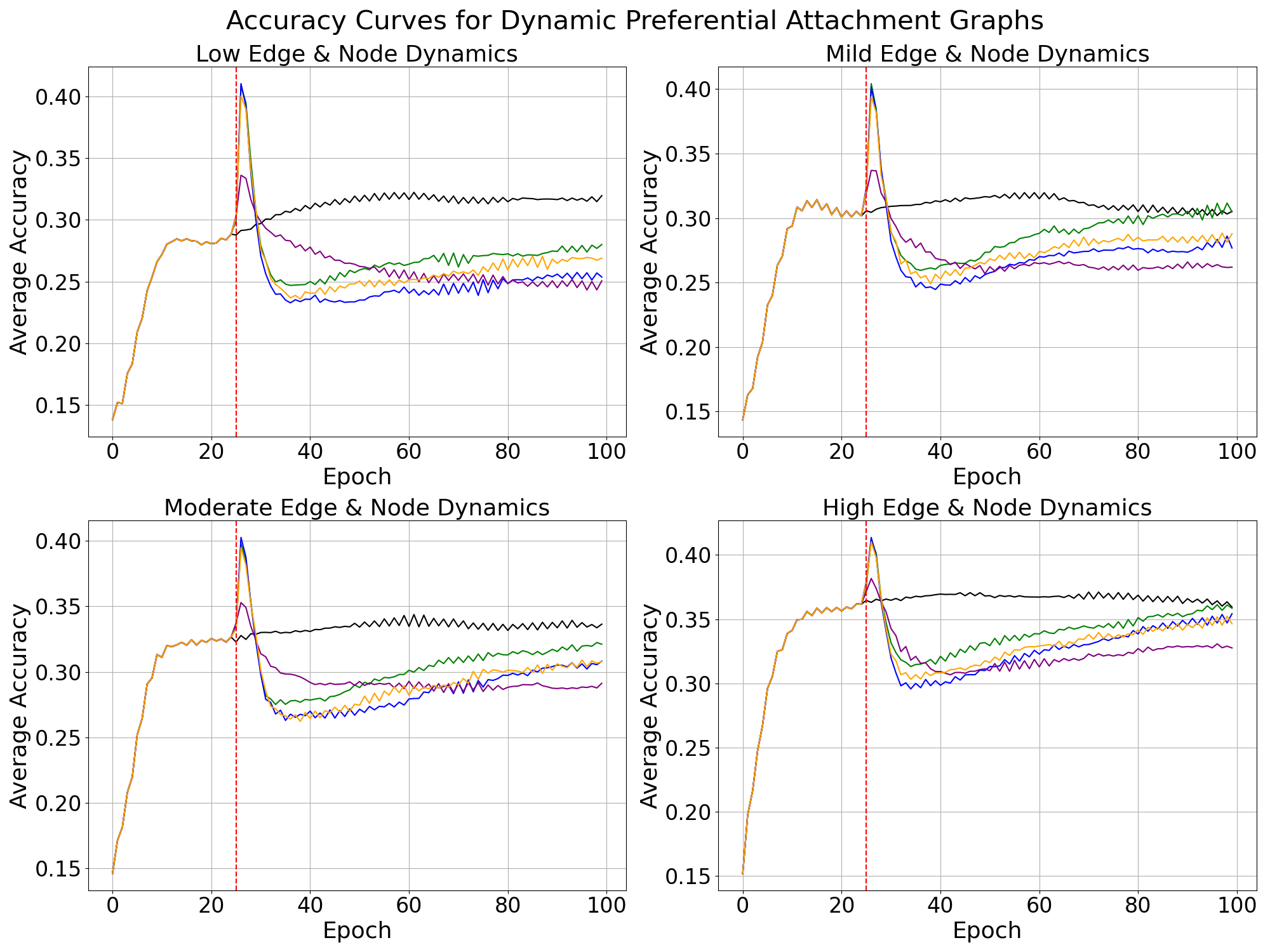}
\end{subfigure}%
\vspace{0.01em}
\caption{\rv{Average Testing accuracy over time of honest nodes in 25-node networks for Directed Geometric graphs with $r = 0.2$ and Preferential Attachment graphs with starting node number $= 1$ and IID data distribution, under link and node failure conditions.}}
\label{fig:dynamic_graphs}
\vspace{-2em}
\end{figure}

For both the DG and PA graphs there are clear differences in behavior as compared to the static graph scenario.

In the DG graph case, both the MaxSpAN-FL and Hopping-Augmented MaxSpAN-FL remain as the two best attacks for all configurations, however both the accuracy decrease and the performance difference diminish as the graph dynamics increase. In the low dynamics setting, the final accuracy achieved by Hopping-Augmented MaxSpAN-FL is $21.9\%$, while in the high dynamics setting this goes up to $27.6\%$. Furthermore in the moderate setting the regular MaxSpAN-FL has a final accuracy of $24.2\%$, which is a noticeably better result than the Hopping-Augmented MaxSpAN-FL, that achieves $25.3\%$. While this difference is small, it's not insignificant. This performance drop is likely caused by the combination of changing network dynamics. All proposed attacks are based on the principle of choosing "the best" adversarial node, according to the attack, given the existing graph structure. With links and nodes failing, the graphs structure changes, and as a result, the new "best" nodes could differ from the ones chosen initially. Despite that, both Hopping-Augmented and regular MaxSpAN-FL attacks maintain their performance edge over other attack schemes, which demonstrates some degree of robustness against changing network dynamics. 

In the PA graph case, the behavior is different. Not only are the attacks less robust to the network changes, but in addition the  random attack appears to perform best in the mild, moderate, and high dynamics settings, in regards to the final accuracy. Even though both MaxSpAN-FL and Hopping-Augmented MaxSpAN-FL reach their lowest accuracy point faster than the random attack, their accuracy actually improves over time, leading to random achieving lower final accuracies in all cases. This is most likely caused by the specific structure of the PA graph. Because this graph has only 1 starting node it is a sparse graph. As a result, even small changes in node and edge dynamics may have great impact on the graph topology. Removing enough edges and nodes may easily lead to the creation of disjoint subgraphs, significantly altering the adversarial behavior. Despite this, the speed at which MaxSpan-FL and Hopping-Augmented MaxSpAN-FL decrease the accuracy initially is remains superior to any other attack scheme, which suggests some degree of robustness to network changes in PA graphs. 

While these experiments only evaluate one simplified scenario, they demonstrate that the proposed attacks achieve good robustness on DG graphs, and some robustness on PA graphs, while confirming that the overall attack performance deteriorates as the network evolves from the state that was used to choose the adversarial nodes. Also of note is the fact that in real-world deployment, majority of wireless networks are distributed and more akin to the geometric networks in which our proposed attacks demonstrated good robustness.

}

\subsubsection{Effects of \rv{Changing Attack Characteristics}} \label{sec:CAC}

\begin{figure}[ht]
\centering

\begin{subfigure}{\columnwidth}
  \centering
  \includegraphics[width=\linewidth]{pictures/timing_attack_legend.png}
\end{subfigure}%
\vspace{-0.15em}fi
\begin{subfigure}{\columnwidth}
  \centering
  \includegraphics[width=\linewidth]{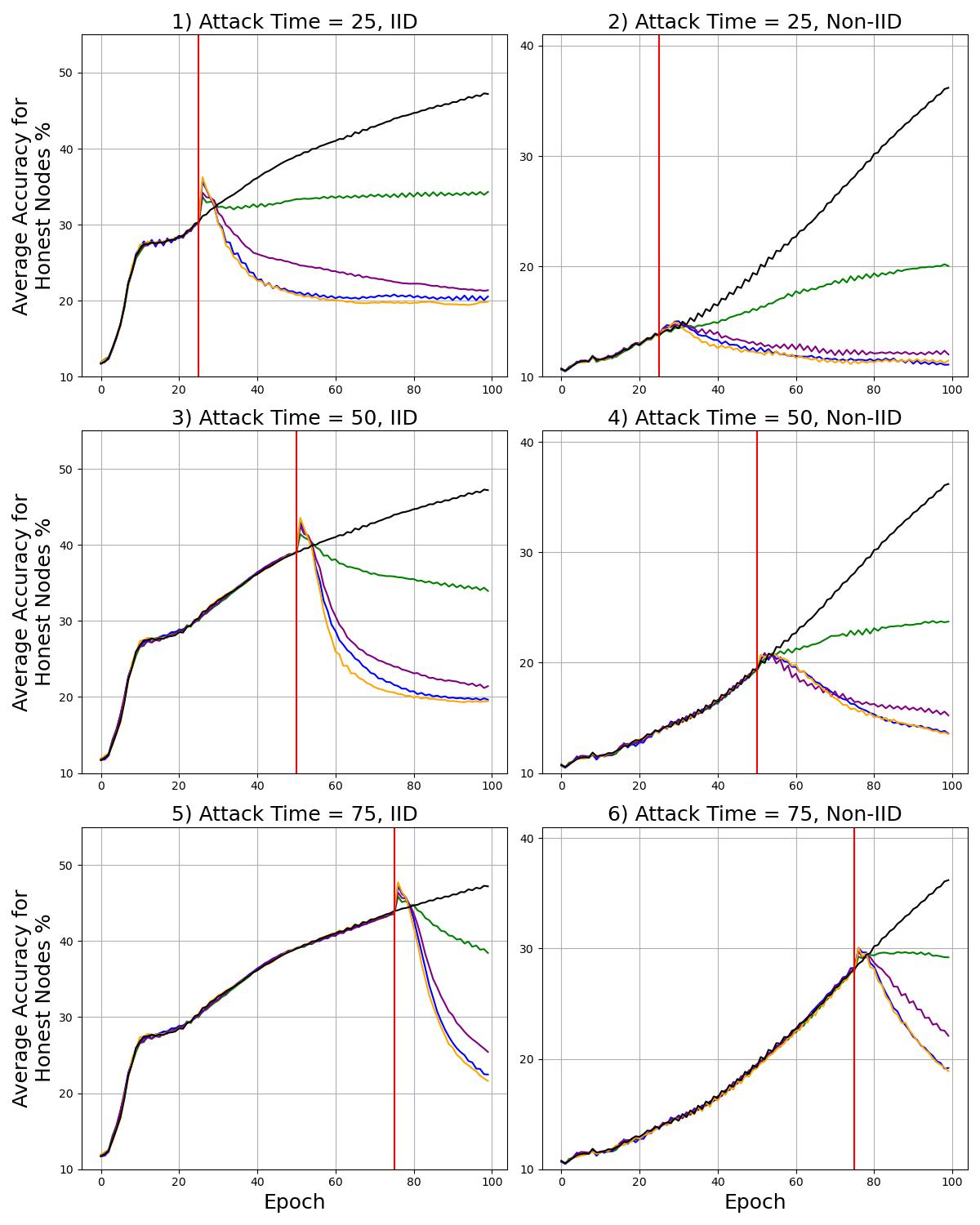}
\end{subfigure}%
\vspace{-0.75em}
\caption{Average testing accuracy over honest nodes in 25-node Directed Geometric Graph with $r=0.2$, for both IID and Non-IID data distributions for different attack deployment times. Adversarial percentage is $20\%$.}
\label{fig:timing_dir_geom}
\vspace{-0.5em}
\end{figure}

\rv{Graph properties such as connectivity, size, or the dynamics of node and edge changes constitute a large part of the properties that influence the analyzed model. However, another significant influence on the environment is created by the characteristics of the actual poisoning attack deployed in the adversarial nodes. We therefore perform experiments on various poisoning attack factors to evaluate their impact on the performance of proposed schemes. 

We first conduct the experiments to investigate} the impact of varying attack deployment times during global aggregation. We focus on the Directed Geometric (DG) graph with 25 clients and radius \( r = 0.2 \), as in this case we evaluate how model optimality affects the attacks difference, which should be independent of the network type. The findings are depicted in Figure \ref{fig:timing_dir_geom}.

Across various attack timings, both the Hopping-Augmented MaxSpAN-FL and the regular MaxSpAN-FL attacks consistently demonstrate superior performance in terms of both final model accuracy and adversarial accuracy loss. This suggests that the effectiveness of the proposed attack is largely unaffected by its deployment time. Notably, in Non-IID scenarios, the  final accuracy discrepancy widens with later attack times. This trend could intuitively be attributed to the degree of model
convergence at the point of attack, although this requires further investigation. Additionally, while attacks initiated at later stages do not reach convergence within 100 epochs, their trajectories seem to be aligning towards final accuracy levels similar to those observed in the 25-epoch attack scenario.

\rv{
Next, we analyze the effects of varying attack powers. The results of those experiments for 3 different graph types are depicted in Figure \ref{fig:attack_strengths}. The attack power in this context refers to the scaling variable $\epsilon_i$ in equation \ref{eq:fgsm_eq} 
, and is set to 50, 100, 250, 500, and 1000 in this experiment. 

\begin{figure}[t]
\centering
\begin{subfigure}{\columnwidth}
  \centering
  \includegraphics[width=\linewidth]{pictures/timing_attack_legend.png}
\end{subfigure}%
\vspace{-0.15em}
\begin{subfigure}{\columnwidth}
  \centering
  \includegraphics[width=\linewidth]{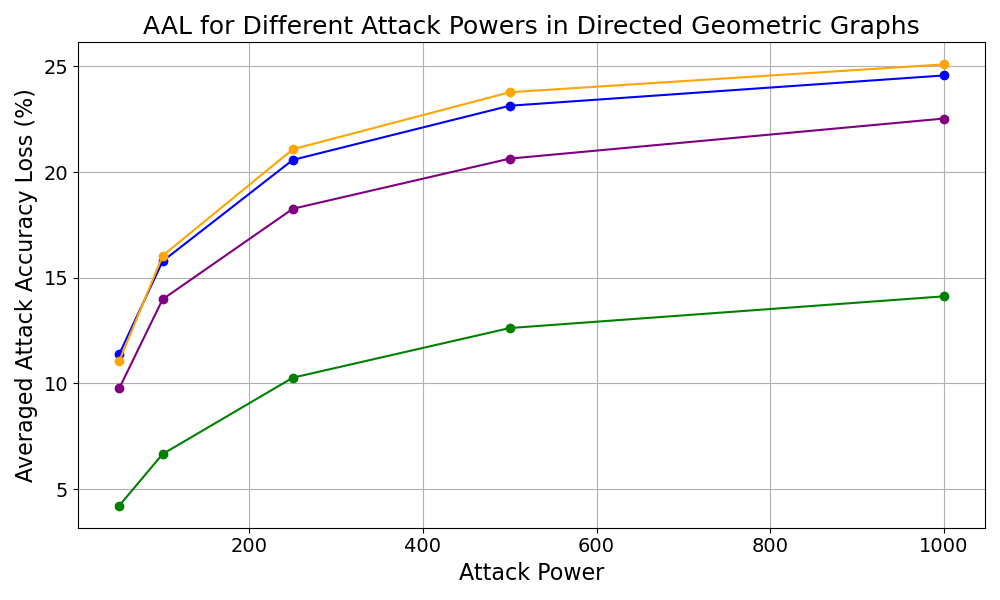}
\end{subfigure}%
\vspace{-0.15em}
\begin{subfigure}{\columnwidth}
  \centering
  \includegraphics[width=\linewidth]{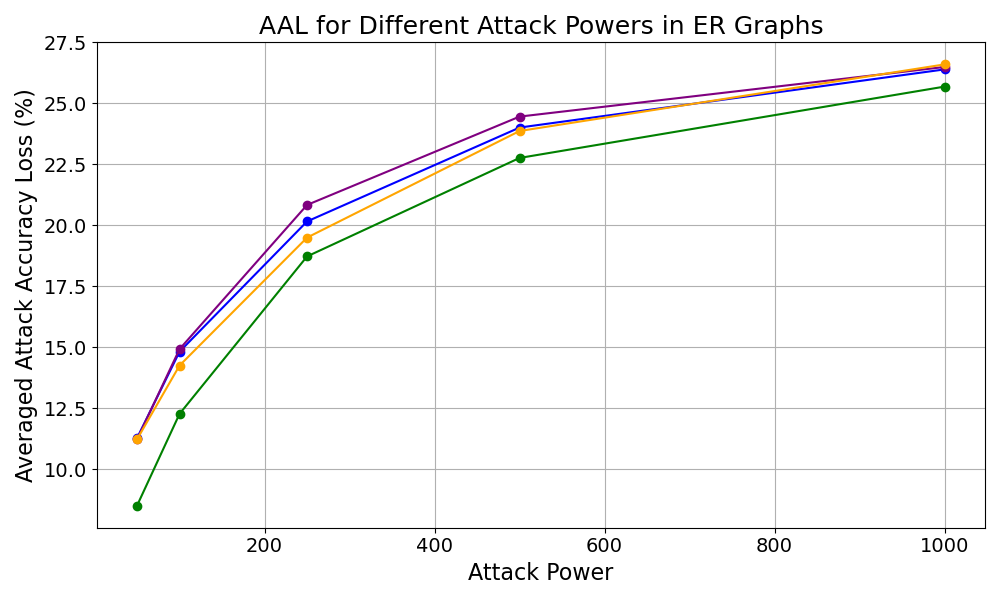}
\end{subfigure}%
\vspace{-0.15em}
\begin{subfigure}{\columnwidth}
  \centering
  \includegraphics[width=\linewidth]{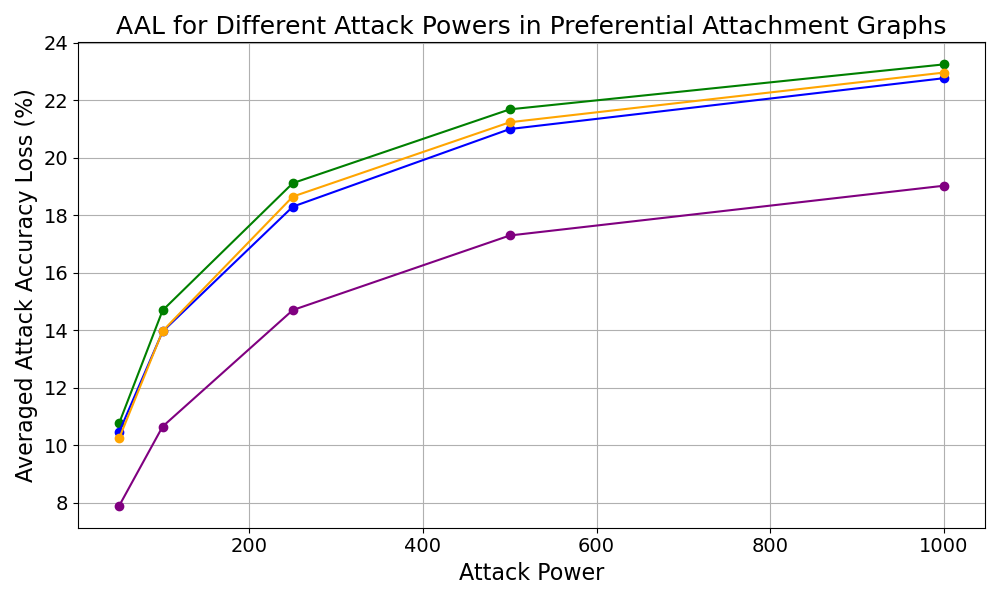}
\end{subfigure}
\vspace{-0.5em}
\caption{\rv{Average Attack Accuracy Loss (AAL) for Directed Geometric graphs with $r = 0.2$, ER graphs with $p = 0.3$ and Preferential Attachment graphs with starting node number $= 1$ and IID data distribution, for different attack power values. Higher AAL corresponds to a more effective attack.}}
\label{fig:attack_strengths}
\vspace{-2em}
\end{figure}

In the 3 tested graphs, the relationships between attack performance are preserved regardless of the power behind the attack, which may indicate that for a given graph the best attack remains constant for all different attack power values. As expected, the Hopping-Augmented MaxSpAN-FL and regular MaxSpAN-FL perform best in the Directed Geometric graph, outperforming the random attack by a $11.37\%$ at the power of 1000. On the ER and Preferential Attachment graphs, both Hopping-Augmented and regular MaxSpAN-FL achieve competitive performances, with the better attack always being within $1\%$ AAL to the best attack in the given scenario. 

In addition to demonstrating good performance of the proposed attacks, this experiment confirms several expected behaviors. Firstly, for all attack powers, the proposed attacks perform similarly well to the random attack, with the performance gap decreasing with increasing attack power, further supporting the hypothesis that node placement isn't as significant of a factor in ER graphs as in other types of graphs. In contrast, in the Preferential Attachment graph all the proposed attacks outperform the baseline random attack for all attack powers by a significant margin, with the Eigenvector-centrality based attack achieving $22.2\%$ attack improvement over random. 

Finally, of note is the relationship between AAL and attack power, present for all tested attacks. The attack power and AAL are proportional to each other, however, as the attack power increases, the rate of increase of AAL is reduced, with the highest rate of change observed between attack powers of 50 and 100, and the lowest between 500 and 1000. This suggest that there is a point of diminishing returns, where further increases in attack power yield progressively smaller gains in AAL.  
There are several factors that could be attributed to causing this behavior. The model exchange mechanism that averages the poisoned models with locally trained models, or the fact that the FGSM attack only poisons the local dataset which means there is a limit on how far the poisoned model can diverge are two most likely explanations for this observed pattern. 

Regardless of the cause, the observed relationship demonstrates that the attack power has significant impact on attack performance, and provides insight in selecting the right parameter. For an attacker this information allows them to select an optimal attack power beyond which not only the additional computation would yield diminishing returns, but also would make the attack easier to detect. By picking a lower optimal power, the attacker can maximize the impact while minimizing resource expenditure and detection probability. 

\subsubsection{Effects of Varying Data Distributions} \label{sec:vdd}
While both the graph and attack properties play significant roles in determining the performance of the proposed adversarial node selection schemes, the distribution of data across nodes remains another crucial factor that can significantly impact the learning process and behavior of adversarial clients in decentralized FL. To analyze this behavior we conduct an experiment with varying levels of data heterogeneity. We vary the number of classes available to each client between 1, 3, 5, 7, and 10, where 10 is the IID case. The results are demonstrated in Figure 
\ref{fig:diff_iid}.

\begin{figure}[t]
\centering
\begin{subfigure}{\columnwidth}
  \centering
  \includegraphics[width=\linewidth]{pictures/legend_diff_graph_sizes.jpg}
\end{subfigure}%
\vspace{-0.15em}
\begin{subfigure}{\columnwidth}
  \centering
  \includegraphics[width=\linewidth]{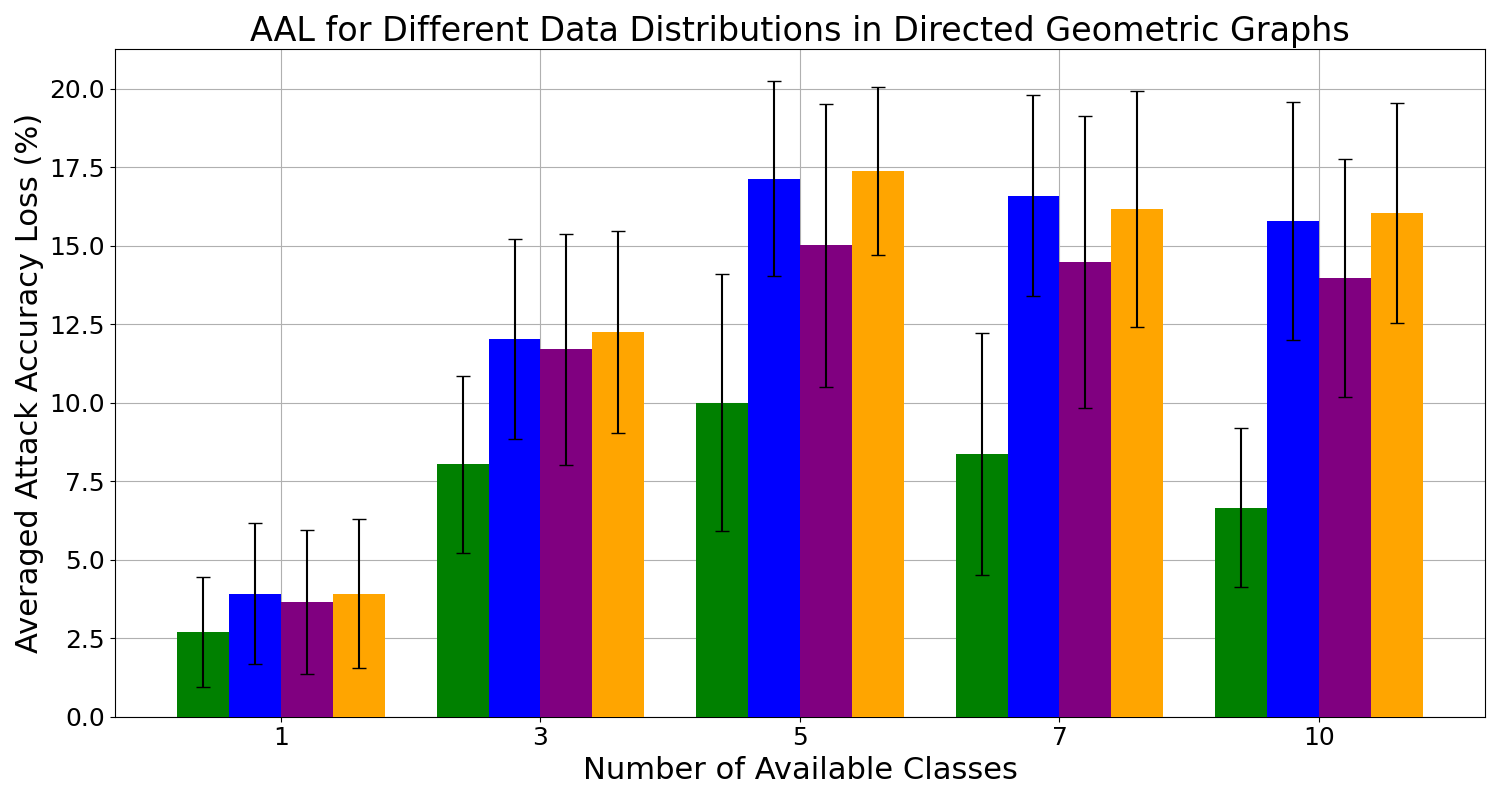}
\end{subfigure}%
\vspace{-0.15em}
\begin{subfigure}{\columnwidth}
  \centering
  \includegraphics[width=\linewidth]{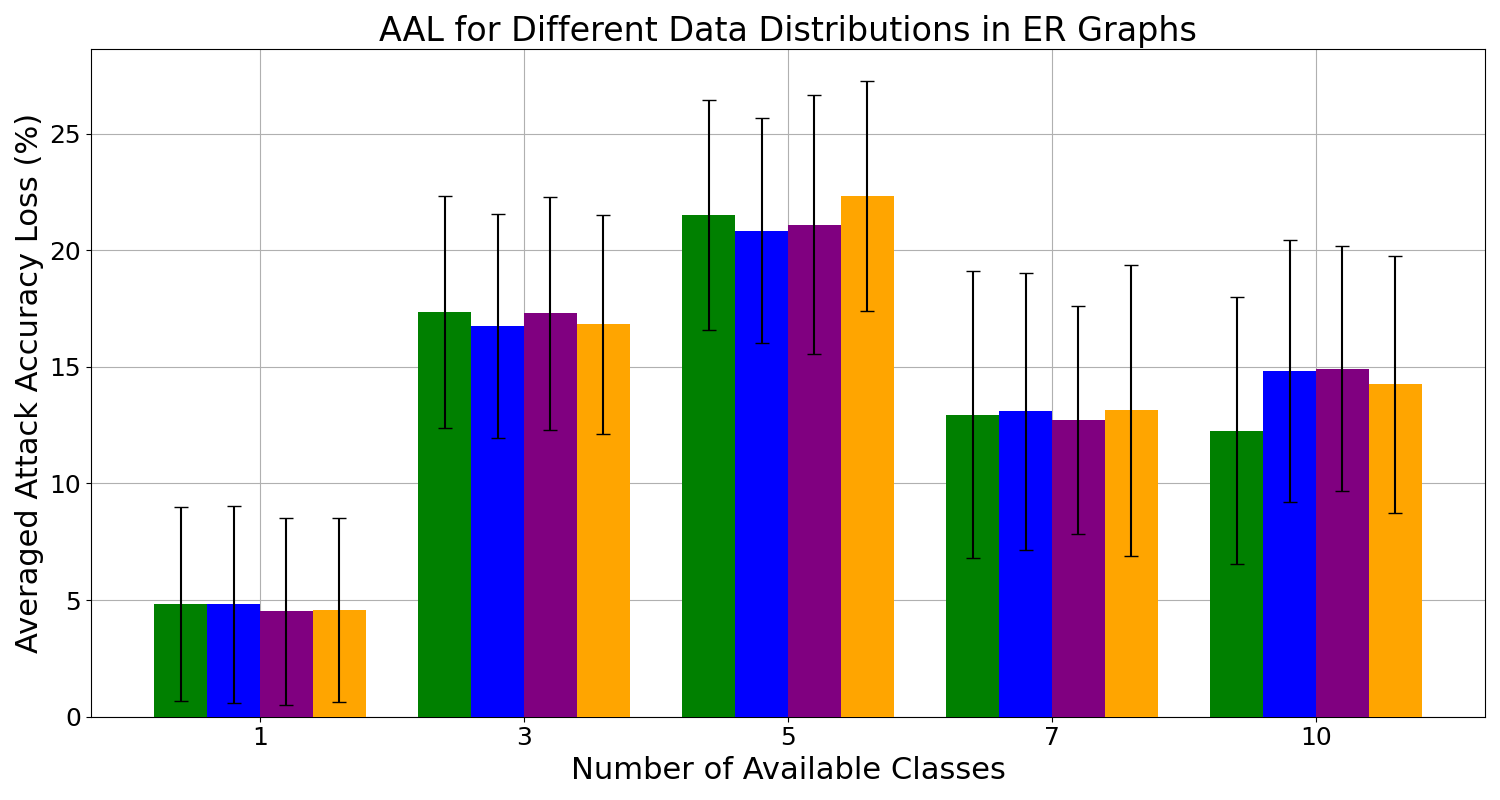}
\end{subfigure}%
\vspace{-0.15em}
\begin{subfigure}{\columnwidth}
  \centering
  \includegraphics[width=\linewidth]{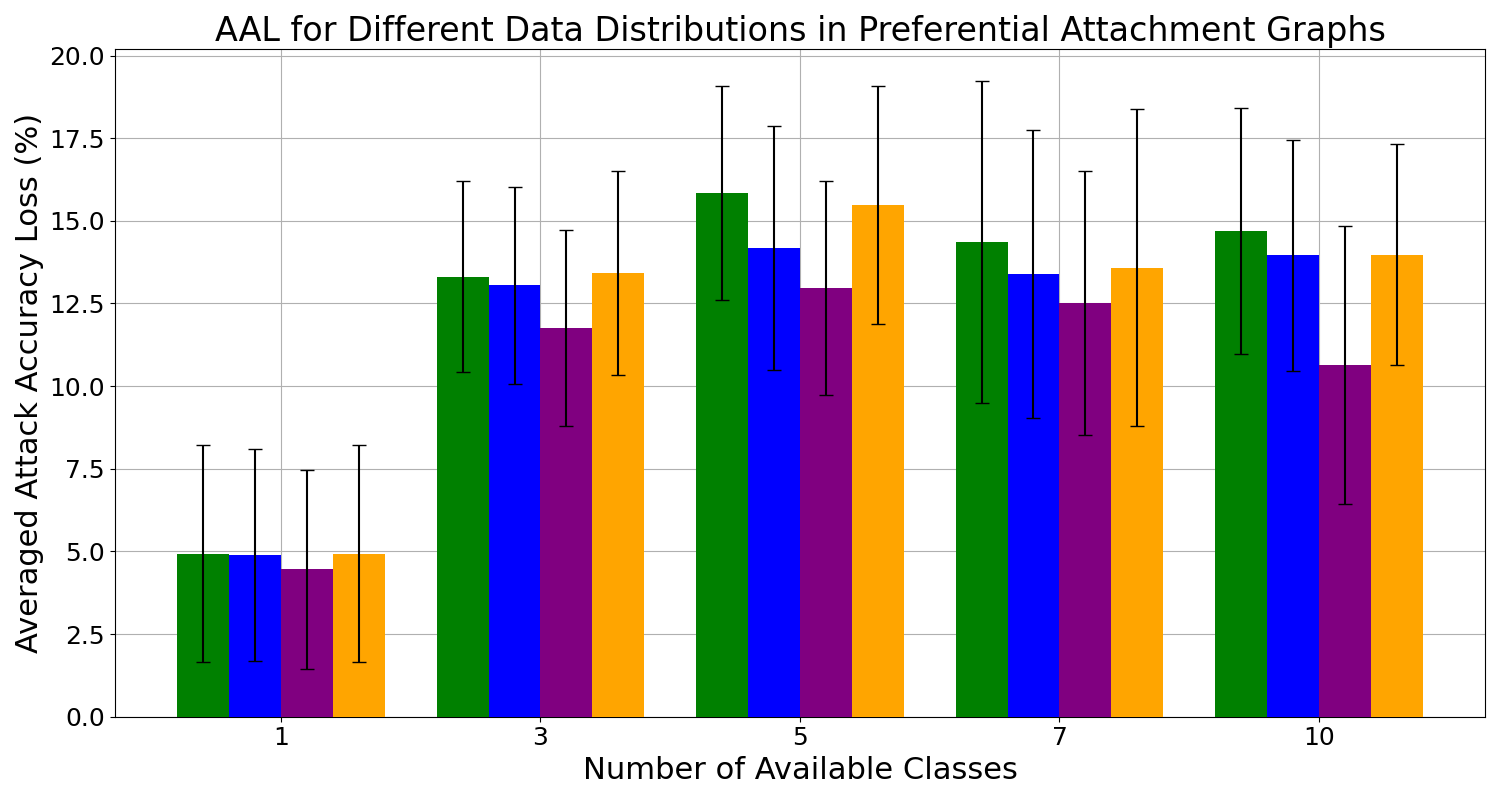}
\end{subfigure}
\vspace{-0.5em}
\caption{\rv{Average Attack Accuracy Loss (AAL) for Directed Geometric graphs with $r = 0.2$, ER graphs with $p = 0.3$ and Preferential Attachment graphs with starting node number $= 1$ and for different data distributions. Higher AAL corresponds to a more effective attack.}}
\label{fig:diff_iid}
\vspace{-2em}
\end{figure}

In the DG graph case, the results demonstrate the efficacy of the MaxSpAN-FL and Hopping-Augmented MaxSpAN-FL, with both attacks achieving best AAL scores for all data distributions. Even though the difference in performance with random is low for the cases where clients had 1 or 3 classes available for training, starting from the case with 5 classes, the two proposed attacks show clear separation in performance to other schemes. The low initial difference for cases with 1 and 3 available classes is explainable by results in Figure \ref{fig:timing_dir_geom}, as for those cases the model has not converged enough to show significant differences when deploying random and MaxSpAN-FL attacks. In fact, in the scenario with 1 available class, the final accuracy after 100 epochs with no attack achieved was $18.7\%$, and all attacks but the Eigenvector-centrality based attack managed to lower the accuracy to the optimum point of approximately $10\%$, at which point the prediction is purely random. Furthermore, in all other cases where the clients had only 1 class available for training, all attacks managed to decrease the accuracy from approximately $20\%$ to $10\%$. 

In the ER graph scenario all attacks perform similarly well, as expected. However, the AAL behavior differs from the DG case. AAL increases for 3 and then for 5 available classes cases, and then decreases for 7 and 10 available classes cases. This however is likely due to the dynamics of the training process. In the ER graph, the final accuracy achieved without any attacks for the is $43.6\%$ in the 3 available classes case, $48.3\%$ in the 5 available classes case, and $49.0\%$ in the 7 available classes case. However, Hopping-augmented MaxSpAN-FL decreases the accuracy to: $10.4\%$, $15.3\%$, and $28.4\%$ for the 3 cases respectively. In the cases with 3 and 5 classes, the final no-attack accuracy is high, but it was lower when the attack began. This also holds for the case with 7 classes, but the accuracy was higher when the attack was deployed compared to other classes. As a result, the attack managed to converge to a lower point for the cases with fewer classes. The reason this only appears in ER graphs is likely due to how well they train the model compared to other types of graphs. The AAL for 7 and 10 available classes is actually comparable to the one achieved in DG or PA graphs, but the AAL for 3 and 5 classes is higher because the final performance of the model is better in ER graphs. ER graphs have on average more edges, so naturally they will have more information exchange and will be more likely to arrive at the optimal model at the end of the training process. 

Finally, in the PA graphs case, the results are again as expected from the previous experiments, and demonstrate that for all cases the Eigenvector-centrality based attack  and the Hopping-Augmented and regular MaxSpAN-FL outperform the random baseline attack. Furthermore, in all cases either the Hopping-Augmented MaxSpAN-FL or 
 the regular MaxSpAN-FL attacks closely match the performance of the Eigenvector-centrality based attack.

}
\rv{\subsection{Limitations} \label{sec:limitations}

The above experiments illustrate a clear relationship between attack strategy and testing accuracy in various FL scenarios. However, several important questions remain unresolved. We provide a brief overview these questions and discuss the limitations of our experimental framework.  

Several open questions are related to the role of the testing data distribution in this relationship. Overall, the observed trends where clearer for IID data distributions compared to Non-IID data distributions. This is evident in both Figure \ref{fig:main_result} and Figure \ref{fig:connectivity_results}, where greater variation in both accuracy and AAL is observed over all attack strategies for the the IID setting compared to the Non-IID setting. While we evaluated different degrees of data heterogeneity, the results revealed trends whose cause we can only theorize. To better understand the impact of adversarial placement in the Non-IID setting, further experiments with a larger number of trials and a variety of data and graph distributions are needed.

Furthermore, the importance of network centrality in adversarial placement is still not completely clear. While our experiments demonstrate that both the vanilla and augmented MaxSpAN-FL attacks outperform the centrality-based attack when data is IID, the centrality-based attack may still outperform both versions in the Non-IID setting (e.g., see Figure \ref{fig:connectivity_results}: preferential attachment graphs). These results indicate that network distribution and data distribution are closely intertwined and cannot be decoupled when trying to understand adversarial placement attacks.

Additionally, the experiments relating to dynamically changing graphs and robustness to link and node failure still leave a lot of room for further exploration. The evaluated model was relatively simplistic in its approximation of real worlds scenarios, and many settings could still be tested, such as constantly changing or deteriorating graphs. On top of that, there still remain questions about some of the behaviors exhibited by the attacks during the training process with dynamic networks. 

Finally, it remains unclear whether any of the proposed attack strategies come close to achieving optimal performance in any FL scenario. The main difficulty here is showing a non-trivial upper bound on testing accuracy or AAL for a given strategy. Showing this requires a theoretical analysis of the problem and is beyond the scope of the numerical methods used in this Section.}

\rv{
\subsection{Possible Defense Strategies} \label{sec:PDS}
In contrast to the abundance of studies on training defenses in centralized FL setting \cite{Ma2020, Shi2022,Fang2020,Yin2024}, few defenses have been proposed to mitigate data poisoning attacks in the \textit{decentralized} setting considered in our paper. In the latter setting, one proposed method is to detect adversarial users by identifying anomalies in clients’ broadcasted parameters, e.g., \cite{AlMallah2022}. This method involves three steps: 1) at each iteration, client behavior is monitored, 2) adversarial clients are detected and 3) adversaries are removed from the aggregation process. The insights gained in our paper could be used to fine-tune steps 1 and 2. For example, using insights into the worst-case adversarial placement within a network, defenses can be optimized to assign more monitoring and detection resources at network locations where adversaries can do the most damage to training, and fewer resources at other locations.

An alternative approach is to modify graph characteristics to ensure better attack resilience. This novel idea would entail introduction of so called bait nodes into the graph at points that are most likely to be attacked by the proposed adversarial selection algorithms. These nodes would be designed to appear as honest nodes to draw in adversaries but would then not participate in the exchange of information with the other nodes. To determine where to place these nodes one could use MaxSpAN-FL or other proposed attacks to create a bait clone nodes of the nodes most likely to be attacked. Alternatively, one could create a new algorithm that would determine places in the network to insert the bait nodes. By modifying the graph properties to the advantage of honest clients a new potential defense strategy could improve the resilience of the learning process.
}

\subsection{Key Takeaways}
\begin{itemize}
    \item The impact of adversarial node placement is significantly influenced by the network's graph distribution and the data distribution, as shown in Figure \ref{fig:main_result}.
    \item In most scenarios, Hopping-Augmented MaxSpAN-FL and MaxSpAN-FL either surpass or match baseline other method performances. Hopping-Augmented MaxSpAN-FL performance improvement reaches up to around \( 678.6\% \).
    \item Network connectivity plays an important role in determining the effectiveness of different adversary placement strategies and their relative performance. 
    \item Varying graph sizes and \rv{adversarial numbers} affects the performance disparity between adversarial placement strategies. 
    \item The Hopping-Augmented MaxSpAN-FL manages to perform optimally in various network types due to optimizing the choice strategy given its environment. 
    \item \rv{The timing and power of the attack consistently maintain performance difference trend between different placement strategies. A diminishing returns effect is observed with increasing power} 
    \item \rv{The proposed MaxSpAN-FL and Hopping-Augmented MaxSpAN-FL attacs demonstrate robustness in dynamic network scenarios, particularly in DG graphs, where they maintain performance advantages under varying node and link failure conditions.}
    \item \rv{In ER graphs, the performance differences between attack strategies are less pronounced, suggesting that node placement is less critical in such network structures.}
    \item \rv{The effectiveness of different attack strategies varies with data heterogeneity. In extreme Non-IID scenarios (very few classes per client), the performance gap between different attack strategies narrows.}
\end{itemize}

\section{Theoretical Performance}
\label{sec:theory}
\noindent Our MaxSpAN-FL Attack Algorithm blends two different types of strategies for maximizing influence over network topologies: choose nodes that are furthest spread apart (i.e., through our max-span algorithm), and also that are most impactful in their respective neighborhoods (i.e., through our hopping mechanism). Given that influence maximization over graphs is an NP-Hard problem~\cite{panagopoulos2023maximizing}, these schemes are often based on approximations or heuristics without formal optimality guarantees, as is the case in our work. We can, however, develop a formal understanding of our observation from Sec.~\ref{exp_results} that \textit{for various graph types, centrality-based attacks, when used alone, are not optimal for attacking decentralized FL}. For this analysis, we focus on eigenvector-based centrality, given its accessible mathematical properties relative to other centrality measures (furthermore, we also observed low variance between the eigenvector centrality-based attack and the performance of other centrality measures in our setting).

\subsection{Assumptions}
Our analysis will employ the following assumptions in the proof of Lemma \ref{thm:theory}:

\textit{Assumption 1:} The graph $G$ is undirected, i.e., the adjacency matrix $E$ is symmetric.

\textit{Assumption 2:} All nodes have the same degree, i.e., for some integer $d \geq 1$, $d_i \triangleq |j:(j,i) \in E| = d$ for all $i \in V$.

\textit{Assumption 3:} The adversarial clients will always keep their stochastic gradients above a threshold value $\delta_{adv,min}$, in a way that will prevent the honest clients from ever reaching an optimal point.

\subsection{Performance Analysis}
We obtain the following bound on the change in models introduced by a set of adversaries in the network:

\begin{lemma}[Impact of Adversarial Nodes on Convergence given Node Centralities] \label{thm:theory}
Given a decentralized Federated Learning (FL) network represented by a strongly connected, time-invariant, directed graph $G = (V, E)$, partitioned into a set of adversarial nodes $A$ and honest nodes $H:=V\setminus A$, and under the Assumptions 1 and 2 regarding the undirected graph structure and node degrees, the expected squared Frobenius norm of the difference between the models at time $t+1$ in an adversarial setting $X(t+1)$ and the models in a hypothetical honest setting $\hat{X}(t+1)$ satisfies:
\begin{gather}
    \mathbb{E}\|X(t+1) - \hat{X}(t+1)\|_F^2 \nonumber \\ \geq {\alpha}^2\mathbb{E}||\underset{j \in A}{\sum}v_j\underset{i = 0}{\overset{t}{\sum}}(\delta_{adv,min} - \widetilde{\nabla}\hat{f_j}(x_j^{(i)}))||_2^2 - \nonumber \\{\alpha}^2\mathbb{E}|| \underset{j \in H}{\sum}v_j\underset{i = 0}{\overset{t}{\sum}}(\widetilde{\nabla} f_j(x_j^{(i)}) - \widetilde{\nabla}\hat{f_j}(x_j^{(i)}))||_2^2
\end{gather}
where $\delta_{adv,min}$ is the lower bound on the adversarial gradients, $v_j$ is the eigenvector centrality of node $j$, and $\widetilde{\nabla} f_j$ and $\widetilde{\nabla} \hat{f_j}$ denote the stochastic gradient and its honest counterpart at node $j$, respectively. 
\end{lemma}
This inequality indicates that the presence of adversarial nodes with a minimum attack strength $\delta_{adv,min}$ leads to a deviation from the optimal convergence trajectory. The first term on the right hand side is directly proportional to the eigenvalues of the selected adversarial nodes. However, even with smaller eigenvalues of the remaining honest nodes, the second term can increase. In particular, for each honest node, the eigenvalue is scaled by the difference between the gradient under attack and its honest counterpart, which depends on the device's data distribution and the number of adversaries in its neighborhood. Thus, eigenvector centrality will not generally maximize this lower bound.

\begin{proof}
Given Assumptions 1 and 2, let a given node $v_i \in V$ have a model $x_i^{(t)} \in \mathbb{R}^p$ at time $t \in \mathbb{N}$. We will conduct the analysis assuming each node $i$ uses a standard consensus-only based aggregation algorithm: 
\begin{equation}
    x_i^{(t+1)} = \underset{j \in N_i}{\sum}\frac{x_j^{(t)}}{d_i} - \alpha \widetilde{\nabla} f_i(x_i^{(t)})
\end{equation}
where $N_i$ and $d_i$ are the neighbors and degree of node $i$, respectively. In matrix form, we have that
\begin{equation}
    X(t+1) = MX(t) - \alpha \Delta(t)
\end{equation}
where $X(t) = [x_1^{(t)}, x_2^{(t)}, \cdots, x_{|V|}^{(t)}]^T$, $M$ is a row-stochastic adjacency matrix where the entry $m_{i,j} := \frac{1}{d_i^{in}}$ if $j \in N_i^{in}$ and $0$ otherwise, and $\Delta(t) = [\widetilde{\nabla} f_1(x_1^{(t)}), \widetilde{\nabla} f_2(x_2^{(t)}), \cdots, \widetilde{\nabla} f_{|V|}(x_{|V|}^{(t)})]^T$.

By Assumption 2, $M = \frac{1}{d} E$, and, thus, the right principal eigenvector $\mathbf{v}$ of $E$ (i.e., the eigenvector-centrality vector) is also the right principal eigenvector of $M$. Furthermore, by Assumption 1, $\mathbf{v}$ is also the left eigenvector of $M$. It follows that
\begin{gather}
    \mathbf{v}^TX(t+1) = \mathbf{v}^T(AX(t) - \alpha\Delta(t)) \nonumber \\
    =  \mathbf{v}^TAX(t) - \alpha \mathbf{v}^T\Delta(t) = \mathbf{v}^TX(t) - \alpha \mathbf{v}^T\Delta(t)
\end{gather}
In turn, by the Perron-Frobenius theorem, the largest unique eigenvalue of $M$ is $1$, and, thus, 
\begin{equation}\label{eig_cent_eq}
    \mathbf{v}^TX(t+1) = \mathbf{v}^TX(t) - \alpha \mathbf{v}^T\Delta(t).
\end{equation}

Next, for our analysis we introduce $X^*$, which contains optimal models for each node, and $\hat{X}(t)$, which is the equivalent of $X(t)$ in a situation where every node in the network is an honest node. 

We will consider the value of optimality gap $(X(t) - X^*) - (\hat{X}(t) - X^*) = X(t) - \hat{X}(t)$ in our analysis. It compares the difference between the node models in the presence of adversaries to the node models when all nodes are honest.


Consider the value of $\mathbb{E}||X(t+1) - \hat{X}(t+1)||_F^2$. Given this setup, it follows that:
\begin{gather*}
    \mathbb{E}||(X(t+1) - \hat{X}(t+1))||_F^2  \\ \geq \underset{\mathbf{w}:\|\mathbf{w}\|_2 = 1}{\max}
    \mathbb{E}||\mathbf{w}^T(X(t+1) - \hat{X}(t+1))||_2^2  \\ \geq
     \mathbb{E}||\mathbf{v}^T(X(t+1) - \hat{X}(t+1))||_2^2\\ 
    = \mathbb{E}||\mathbf{v}^T((AX(t) - \alpha \Delta(t)) - (A\hat{X}(t) - \alpha \hat{\Delta}(t)))||_2^2 \\
    = \mathbb{E}||\mathbf{v}^T((AX(t) - A\hat{X}(t)) - \mathbf{v}^T\alpha (\Delta(t) - \hat{\Delta}(t))||_2^2 \\
    \overset{(a)}{=} \mathbb{E}||\mathbf{v}^T((X(t) - \hat{X}(t)) - \mathbf{v}^T\alpha (\Delta(t) - \hat{\Delta}(t))||_2^2 \\
    =  \mathbb{E}||\mathbf{v}^T((X(0) - \hat{X}(0)) - \underset{i = 0}{\overset{t}{\sum}}\mathbf{v}^T\alpha (\Delta(i) - \hat{\Delta}(i))||_2^2 \\
    \overset{(b)}{=} \mathbb{E}||\underset{i = 0}{\overset{t}{\sum}}\mathbf{v}^T\alpha (\Delta(i) - \hat{\Delta}(i))||_2^2 \\
    = {\alpha}^2\mathbb{E}||\mathbf{v}^T \underset{i = 0}{\overset{t}{\sum}}(\Delta(i) - \hat{\Delta}(i))||_2^2 \\
    = {\alpha}^2\mathbb{E}||\underset{j \in A}{\sum}v_j\underset{i = 0}{\overset{t}{\sum}}(\widetilde{\nabla} f_j(x_j^{(i)}) - \widetilde{\nabla}\hat{f_j}(x_j^{(i)})) + \\ \underset{j \in H}{\sum}v_j\underset{i = 0}{\overset{t}{\sum}}(\widetilde{\nabla} f_j(x_j^{(i)}) - \widetilde{\nabla}\hat{f_j}(x_j^{(i)}))||_2^2 \\
    \overset{(c)}{=}  {\alpha}^2\mathbb{E}||\underset{j \in A}{\sum}v_j\underset{i = 0}{\overset{t}{\sum}}(\delta_{adv,min} - \widetilde{\nabla}\hat{f_j}(x_j^{(i)})) + \\ \underset{j \in H}{\sum}v_j\underset{i = 0}{\overset{t}{\sum}}(\widetilde{\nabla} f_j(x_j^{(i)}) - \widetilde{\nabla}\hat{f_j}(x_j^{(i)}))||_2^2 \\
    \geq {\alpha}^2\mathbb{E}||\underset{j \in A}{\sum}v_j\underset{i = 0}{\overset{t}{\sum}}(\delta_{adv,min} - \widetilde{\nabla}\hat{f_j}(x_j^{(i)}))||_2^2 - \\{\alpha}^2\mathbb{E}|| \underset{j \in H}{\sum}v_j\underset{i = 0}{\overset{t}{\sum}}(\widetilde{\nabla} f_j(x_j^{(i)}) - \widetilde{\nabla}\hat{f_j}(x_j^{(i)}))||_2^2
\end{gather*}
where (a) follows from \ref{eig_cent_eq}, (b) follows from the assumption that the initial models for cases with and without malicious agents are the same, and (c) follows from Assumption 3.
\end{proof}

\rv{\subsection{Complexity Analysis}\label{sec:complexity_analysis}

Recall that there are $n$ nodes in the network graph $G$ and let the number of adversarial nodes $n_{\mathrm{adv}}$ be a function of $n$. Assume that the limit $\lim_{n \rightarrow \infty} n_{\mathrm{adv}}/n$ exists and is strictly less than $1$. In this Section, we show that the time complexity of the MaxSpAN-FL algorithm is $O(n^3)$. Recall that MaxSpAN-FL chooses a fixed-sized subset $A \subset \{1,2,\ldots,n \}$ of adversarial nodes via the following two step procedure. 

 In the first step, for each node $g \in G$, MaxSpAN-FL computes an influence region $C[g] \subset \{1,2,\ldots,n\}$ of size $S_{\mathrm{cluster}}$ by traversing the $S_{\mathrm{cluster}}$ closest nodes to $g$ using breadth-first-search (BFS). For a fixed $g \in G$, the time to run BFS is $O(S_{\mathrm{cluster}} + E_{\mathrm{cluster}})$ where $E_{\mathrm{cluster}}$ is the number of edges in the subgraph consisting of nodes $C[g]$. Thus, the time to run the first step is 
$O\left(n \left(S_{\mathrm{cluster}} + E_{\mathrm{cluster}}\right)\right)$. This is at most $O\left(n\left(S_{\mathrm{cluster}} + S^2_{\mathrm{cluster}}\right)\right) = O(n^3)$ following that (a) the number of edges $E_{\mathrm{cluster}}$ of the directed subgraph is at most $ S_{\mathrm{cluster}} (S_{\mathrm{cluster}}-1)$ and (b) $S_{\mathrm{cluster}} \approx n/n_{\mathrm{adv}} = O(n)$.

 In the second step, MaxSpAN-FL chooses the subset $A$ that approximately minimizes the overlap between influence regions. For fixed $A$ and $H$ and for $g \in H$, the time to compute overlap $o[g] \triangleq |C[g] \cap (\cup_{g' \in A} C[g'])|$ is at most $O(n)$ as the time to compute the intersection is $O(n)$. Thus, computing overlaps for all $|H| \in \{n,n-1,\ldots, n-n_{\mathrm{adv}}+1$ \} takes 
\begin{align}
&O( n \cdot n + (n-1)n + (n-2)n + \ldots + (n-n_{\mathrm{adv}}-1)n) \nonumber \\  &= O((n^2 - n^2_{\mathrm{adv}})n) \nonumber = O(n^3) \nonumber
\end{align}
time, where the last equality follows from the assumption that $\lim_{n \rightarrow \infty} n_{\mathrm{adv}}/ n < 1$.

}

\section{Conclusion and Future Work}
\label{sec:conclusion}
\noindent The rapid advancement of decentralized federated learning (FL) has introduced a range of novel applications, while simultaneously presenting unique security challenges. In this paper, we delved into the impact of adversarial node placement within decentralized FL environments, a critical aspect often influenced by the network topology. Our investigations revealed that the effect of adversarial node placement on attack potency in decentralized FL depends on a variety of elements, including network type, connectivity, size, \rv{dynamics}, the timing \rv{and power} of the attack, \rv{and the data distribution}. We also developed a novel attack placement algorithm that blends a maximum distance approach with a centrality-based scheme, and demonstrated superior performance in attack potency across diverse conditions. Finally, we presented a theoretical argument as to why eigenvector centrality alone is not sufficient to optimize the attack impact on decentralized FL \rv{as well as formalized the complexity analysis for the proposed new attack algorithm}.

In future work, we plan to more carefully consider the role of data heterogeneity in optimizing attack strategies, as well as potential defenses against node-selection attack frameworks.


\bibliography{ref.bib}
\bibliographystyle{IEEEtran}

\end{document}